\documentclass[fleqn,usenatbib]{mnras}

\usepackage{newtxtext,newtxmath}
\usepackage{soul}

\usepackage[T1]{fontenc}

\DeclareRobustCommand{\VAN}[3]{#2}
\let\VANthebibliography\thebibliography
\def\thebibliography{\DeclareRobustCommand{\VAN}[3]{##3}\VANthebibliography}

\usepackage{graphicx}	
\usepackage{amsmath}	

\title[The $z=9-15$ UV luminosity function]{JWST PRIMER: A new multi-field determination of the evolving galaxy UV luminosity function at redshifts $\mathbf{z \simeq 9-15}$}

\author[C.\,T. Donnan et al.]{C.\,T. Donnan$^{1}$\thanks{E-mail: callum.donnan@ed.ac.uk},
R.\,J. McLure$^{1}$,
J.\,S. Dunlop$^{1}$,
D.\,J. McLeod$^{1}$,
D. Magee$^{2}$,
K.\,Z. Arellano-C\'{o}rdova$^{1}$, \and
L. Barrufet$^{1}$,
R. Begley$^{1}$,
R.\,A.\,A, Bowler$^{3}$,
A.\,C. Carnall$^{1}$,
F. Cullen$^{1}$,
R.\,S. Ellis$^{4}$,
A. Fontana$^{5}$, \and
G.\,D. Illingworth$^{2}$, 
N.\,A. Grogin$^{6}$,  
M.\,L. Hamadouche$^{1}$,
A.\,M. Koekemoer$^{6}$,
F.-Y. Liu$^{1}$,
C. Mason$^{7,8}$ \and
P. Santini$^{5}$, 
T.\,M. Stanton$^{1}$
\\
$^{1}$Institute for Astronomy, University of Edinburgh, Royal Observatory, Edinburgh, EH9 3HJ, UK\\
$^{2}$Department of Astronomy and Astrophysics, UCO/Lick Observatory, University of California, Santa Cruz, CA 95064, USA\\
$^{3}$Jodrell Bank Centre for Astrophysics, Department of Physics and Astronomy, School of Natural Sciences, The University of Manchester, Manchester, M13 9PL, UK\\
$^{4}$Department of Physics \& Astronomy, University College London. Gower St., London WC1E 6BT, UK\\
$^{5}$INAF - Osservatorio Astronomico di Roma, via di Frascati 33, 00078 Monte Porzio Catone, Italy\\
$^{6}$Space Telescope Science Institute, 3700 San Martin Drive, Baltimore, MD 21218, USA\\
$^{7}$Cosmic Dawn Center (DAWN), Jagtvej 128, DK-2200, Copenhagen N, Denmark\\
$^{8}$Niels Bohr Institute, University of Copenhagen, Jagtvej 128, DK-2200, Copenhagen N, Denmark\\
}

\date{Accepted XXX. Received YYY; in original form ZZZ}

\pubyear{2015}

\begin{document}
\label{firstpage}
\pagerange{\pageref{firstpage}--\pageref{lastpage}}
\maketitle

\begin{abstract}
We present a new determination of the evolving galaxy UV luminosity function (LF) over the redshift range $8.5<z<15.5$ using a combination of several major Cycle-1 {\it JWST} imaging programmes - PRIMER, JADES and NGDEEP. This multi-field approach yields a total of  $\simeq370$ sq. arcmin of \textit{JWST}/NIRCam imaging, reaching (5-$\sigma$) depths of $\simeq30$ AB mag in the deepest regions. We select a sample of 2548 galaxies with a significant probability of lying at high redshift ($p(z>8.5)>0.05$) to undertake a statistical calculation of the UV LF. Our new measurements span $\simeq4$ magnitudes in UV luminosity at $z=9-12.5$, placing new constraints on both the shape and evolution of the LF at early times. Our measurements yield a new estimate of the early evolution of cosmic star-formation rate density ($\rho_{\rm{SFR}}$) confirming the gradual decline deduced from early \textit{JWST} studies, at least out to $z \simeq 12$. Finally we show that the observed early evolution of the galaxy UV LF (and $\rho_{\rm{SFR}}$) can be reproduced in a ${\rm \Lambda}$CDM Universe, with no change in dust properties or star-formation efficiency required out to $z \simeq 12$. Instead, a progressive trend towards younger stellar population ages can reproduce the observations, and the typical ages required at $z \simeq$ 8, 9, 10, and 11 all converge on $\simeq 380-330$\,Myr after the Big Bang, indicative of a rapid emergence of early galaxies at $z \simeq 12 - 13$. This is consistent with the first indications of a steeper drop-off in $\rho_{\rm{SFR}}$ we find beyond $z \simeq 13$, possibly reflecting the rapid evolution of the halo mass function at earlier times. 
\end{abstract}

\begin{keywords}
galaxies:high-redshift -- galaxies:evolution -- galaxies:formation
\end{keywords}



\section{Introduction}
The discovery and study of the earliest galaxies is critical for refining our understanding of cosmology and structure formation, and in particular for clarifying the first stages of galaxy formation/evolution and the progress/drivers of cosmic hydrogen reionization \citep{dunlop2013,stark2016}. Over the last decade, deep near-infrared surveys with the \textit{Hubble Space Telescope} ({\em HST}) and \textit{Spitzer} have enabled the evolution of galaxies to be mapped out to redshifts $z\simeq 9$ \citep[e.g.][]{ellis2013, mclure2013, finkelstein2015, mcleod2015, mcleod2016,oesch2018, bouwens2021, bouwens2022}. Moreover, there is now growing evidence that the UV radiation emitted from these early star-forming galaxies powered cosmic hydrogen reionization, with this phase transition concluding by $z\simeq5.5-6.0$ \citep{robertson2015,aird2015,bosman2021}.

The reliable detection of galaxies at $z\geq7$ was first made possible by the installation of the near-infrared camera WFC3/IR on {\em HST}. The detection of galaxies at $z\geq6$ relies on the robust identification of the Lyman-break due to the attenuation of UV photons short-ward of $\lambda_{\rm{rest}}= 1216$\,\AA\ produced by the  increasingly neutral intergalactic medium (IGM) at these redshifts. However, as a result of the long-wavelength limit of {\it HST} at $1.7 \mu$m, galaxies could only be robustly selected up to $z\sim10$ with only a small number of more tentative detections reported at higher redshifts \citep[e.g.][]{ellis2013,oesch2016} which relied on single filter detections. In parallel with these space-based efforts, wider area (degree-scale)  ground-based near-infrared imaging surveys \citep[e.g. UltraVISTA;][]{mccracken2012} have revealed an excess of bright galaxies ($M_{\rm{UV}} \leq -20$) at high redshifts compared to what is expected from an evolving  Schechter function \citep[][]{bowler2020,varadaraj2023,donnan2023a}. When the ground-based and space-based data are combined it is now clear that a double-power law is a more appropriate functional form for describing the UV luminosity function (LF) at $z>6$. 

Now, since summer 2022, the study of early galaxies has been revolutionised by the advent of {\it JWST}. The NIRCam instrument on {\it JWST} provides deep multi-band near-infrared imaging out to $\lambda \simeq 5 \mu$m, and this has enabled  the robust detection of galaxies at extreme redshifts ($z\geq10$) for the first time. Indeed, in just the first year of {\it JWST} operations, several early NIRCam imaging surveys have already been completed, revealing a significant number of galaxy candidates at $z>10$ \citep[e.g.][]{donnan2023a,donnan2023b,mcleod2023,adams2022b,naidu2022,castellano2022,harikane2023a,finkelstein2022,austin2023,hainline2023}. \textit{JWST} has also been successfully used  to spectroscopically confirm the redshift of many of these candidates using the NIRSpec instrument (up to $z_{\rm{spec}}=13.2$) \citep[e.g.][]{curtislake2022,arrabalharo2023,harikane2023b,wang2023, castellano2024}. These early spectroscopic results have shown very good agreement with the inferred  photometric redshifts for the vast majority of robust targets \citep{arrabalharo2023b,bunker2023}.

Out of the galaxy candidates revealed by {\it JWST} at $z>10$ there have been a number that are particularly (arguably surprisingly) bright \citep[e.g.][]{castellano2022,castellano2023,mcleod2023}. This has led to some suggestions that the early {\it JWST} results in this field present a challenge theoretical models of galaxy evolution, with lower dust attenuation and/or higher star-formation efficiencies, or even primeval (PopIII) stellar populations  being proposed as necessary to explain the observations \citep[e.g.][]{tacchella2018,yung2019,mason2022,harikane2023a}.

Prior to the launch of \textit{JWST}, and due primarily to the aforementioned  wavelength limitations of {\it HST}, there was uncertainty over the evolution of the cosmic star-formation rate density ($\rho_{\rm{SFR}}$) at $z\geq9$. This issue now been largely resolved by the first year of  \textit{JWST} observations,  with early measurements of the UV LF with \textit{JWST} showing only gradual evolution over the redshift range $z=8-13$ \citep[][]{donnan2023a,donnan2023b,harikane2023a,finkelstein2022c,bouwens2023a,bouwens2023b,adams2023,mcleod2023,leung2023}. This is consistent with the smoother more gradual decline in $\rho_{\rm{SFR}}$, consistent with the results/predictions of \cite{mcleod2016}. 

However, the early high-redshift studies undertaken with  \textit{JWST} suffer from several limitations. One limitation is the modest total area coverage and limited availability of deep imaging in the Early Release Science (ERS) and the smaller-scale early Cycle-1 {\it JWST} NIRCam programmes. As a result there has remained significant uncertainty over the exact shape of the UV LF at $z>10$. In particular, the constraints on the faint-end slope of the LF are relatively poor, an uncertainty which in turn limits the accuracy with which the cosmic star-formation rate density can be inferred. Moreover, a free fit of the functional form to the UV LF at $z>10$ has remained challenging due to the limited dynamic range in the UV luminosity ($M_{\rm{UV}}$) of the galaxies uncovered at these redshifts in the early {\it JWST} samples. The limited area of these early studies, as well as inevitably yielding rather small samples, also makes them vulnerable to cosmic variance. In particular, the Abell2744 field imaged by the UNCOVER \citep{bezanson2022} and GLASS \citep{treu2022} programmes is now known to be highly over-dense at $z\sim10$ \citep{castellano2023}. 

Another limitation of the early {\it JWST} studies of high-redshift galaxy evolution has been the methodology used to determine the UV LF. Typically, the $1/V_{\textrm{max}}$ method \citep{schmidt1968} has been used, which involves selecting galaxies at high redshift using the Lyman break technique and simply adopting the best-fitting photometric redshift (photo-z). However, the redshift of a galaxy is often not very well defined by the multi-band photometry and therefore the photo-z can be unreliable, particularly for galaxies close to the detection limit. This is one possible explanation for the discrepant galaxy samples produced in a number of the early {\it JWST} studies \citep{bouwens2023b}.

In this paper we aim to substantially advance our knowledge of the evolving galaxy UV LF and hence cosmic star-formation rate density at $z=9-15$. First, we are now able to utilise a number of the more major extragalactic NIRCam imaging surveys (both public and GTO) which have been largely completed in \textit{JWST} Cycle-1. Second, to do this expanded dataset justice, we have implemented  a more statistically robust method of calculating the LF. 

Our primary imaging dataset is the Public Release IMaging for Extragalactic Research (PRIMER; Dunlop et al., in preparation) survey which provides imaging over $\simeq 380\,\rm{arcmin}^2$ in 8 NIRCam filters, an order-of-magnitude larger area than covered in the ERS NIRCam programmes. We also include the ultra-deep imaging from the {\it JWST} Advanced Deep Extragalactic Survey \citep[JADES;][]{eisenstein2023} and the Next Generation Deep Extragalactic Exploratory Public survey \citep[NGDEEP;][]{bagley2023}. By taking this multi-field approach, combining surveys with different (and largely complementary) depths and areas (covering a total of $\sim 400$ sq. arcmin while reaching depths of $m_{\rm{AB}}\simeq30$) we now have substantially improved dynamic range in both UV luminosity and redshift compared to previous \textit{JWST}-based studies. As a result of this, and our improved methodology, we are now able to significantly improve the accuracy with which the UV LF can be constrained at extreme redshifts, $z=8-15$.

The paper is structured as follows. In Section \ref{sec:data} we describe the imaging data and the creation of our source catalogues. In Section \ref{sec:sample} we explain the sample selection and the spectral energy distribution (SED) fitting to the galaxy photometry. In Section \ref{sec:LF} we present our derived galaxy UV luminosity function as a function of redshift, and the resulting inferred early evolution of the cosmic star-formation rate density. In Section \ref{sec:discussion} we then discuss our results in the context of other recent observational studies and the predictions of various theoretical/numerical models of galaxy formation and evolution. Finally, in Section \ref{sec:conclusions} we summarise our conclusions. Throughout we use magnitudes in the AB system \citep{oke1974,oke1983}, and assume a standard cosmological model with $H_0=70$ km s$^{-1}$ Mpc$^{-1}$, $\Omega_m=0.3$ and $\Omega_{\Lambda}=0.7$.

\section{Data}
\label{sec:data}

\subsection{Survey Fields}
We utilise a number of major {\it JWST} Cycle-1 imaging surveys covering 4 separate fields. Due to the uncertainties and limited cosmological volumes associated with gravitational lensing, we choose to focus on blank/unbiased fields to ensure a robust determination of the UV luminosity function. The largest programme used is the Public Release IMaging for Extragalactic Research (PRIMER, PI: J. Dunlop) survey which images the COSMOS and UDS fields using NIRCam through the F090W, F115W, F150W, F200W, F277W, F356W, F410M and F444W filters. We also use the first epoch of imaging from the Next Generation Deep Extragalactic Exploratory Public (NGDEEP, PI: S. Finkelstein) survey which provides ultra-deep imaging in the GOODS-South field, specifically targeting the Hubble Ultra Deep Field parallel 2 field (HUDF par. 2). This programme uses the NIRCam F115W, F150W, F200W, F277W, F356W and F444W filters. Finally we use the {\it JWST} Advanced Deep Extragalactic Survey \citep[JADES;][]{eisenstein2023,rieke2023} NIRCam data release 2. This programme targets a region within the GOODS-South field centred on the Hubble Ultra Deep Field (HUDF), and uses the same filter set as the PRIMER survey. 

These survey fields also have the advantage of deep optical imaging from the \textit{HST}/ACS instrument. The PRIMER UDS and COSMOS fields have been imaged in in the {\it HST} F435W, F606W and F814W filters from the Cosmic Assembly Near-IR Deep Extragalactic Legacy Survey \citep[CANDELS;][]{grogin2011,koekemoer2011}. The JADES and NGDEEP fields also have imaging in these filters with the addition of F775W and F850LP. This imaging was also taken as part of CANDELS as well as through the Great Observatories Origins Deep Survey \citep[GOODS;][]{giavalisco2004}. This includes the deepest \textit{HST}/ACS imaging ever taken in the HUDF. The existence of deep optical data is critical for obtaining robust measurements of the redshift probability distributions ($p(z)$) of the sources detected in each field.

The PRIMER and NGDEEP imaging data was reduced using the PRIMER Enhanced NIRCam Image Processing Library (\textsc{PENCIL}; Magee et al.,  in preparation) software. For the JADES imaging we use the reductions described in \citet{rieke2023}. The astrometry of all the reduced images was aligned to GAIA DR3 \citep{gaia2023} and stacked to the same pixel scale of 0.03\,arcsec.

\subsection{PSF homogenisation}
\label{sec:psf}
To create a catalogue with accurate photometry, the differences in the point spread function (PSF) need to be accounted for. This is corrected by homogenising the PSFs produced by the imaging through the different filters to one common PSF, using a similar technique to that which was utilised for the ground-based imaging in \citet{donnan2023a}. For all of the survey fields used in this study, the F444W imaging is available and has the broadest PSF. The F444W PSF was therefore chosen as the natural target for the homogenisation of the imaging resolution at all other wavelengths. We measured the PSF by selecting $\simeq20$ bright but unsaturated stars across the PRIMER COSMOS imaging. Each star was then centroided and stacked to generate a measurement of the PSF in each filter. Our measured PSFs are comparable to those from WebbPSF \citep{perrin2014} to within $\simeq 1\%-3\%$. Using a combination of Moffat and Gaussian profiles a series of kernels were produced which, when convolved with the imaging, result in the homogenisation of the imaging through each NIRCam filter to the PSF at F444W, to within an accuracy of  3 per cent. 

\subsection{Photometric Depths}
\label{sec:depths}
\begin{figure*}
	\includegraphics[width=\textwidth]{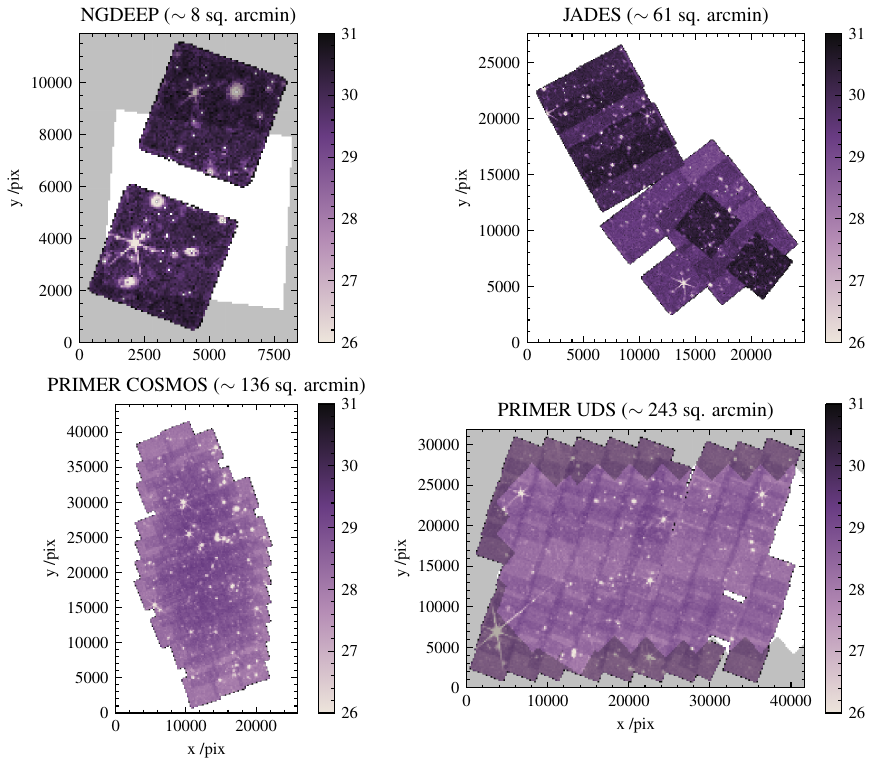}
    \caption{The $5\sigma$ depth maps in the F277W filter imaging of each NIRCam survey field used in this analysis, demonstrating the variation in depths between the different survey fields, and in some cases within a given field. All images are shown with bins of 200 pixels where the original images are on a 0.03-arcsec pixel scale. The colour-bar shows the $5\sigma$ depth in AB mag on the same scale for each field. The grayed out region shows where there is a lack of deep \textit{HST}/ACS F814W imaging, which only affects the NGDEEP and UDS fields.}
    \label{fig:depth_map}
\end{figure*}

The global depths for each field were calculated by determining the distributions of fluxes measured in 0.3-arcsec diameter apertures placed within source-free regions of the PSF-homogenised imaging. The $1\sigma$ depth is then given by $1.4826\times\rm{MAD}$ where MAD is the median absolute deviation of the flux in the source-free apertures. This gives a global depth, because it is calculated over the full (source-free) area of the imaging. In Fig. \ref{fig:depth_map} we show the spatial distribution of the $5\sigma$ depth across each field. This illustrates the different depths achieved by the different NIRCam surveys, as well as revealing the extent of any depth variations across the images. In particular, the ``wedding-cake'' structure of the PRIMER and JADES surveys can be identified by the distinct regions of differing depth. The depth maps indicate that the NGDEEP and UDS fields are relatively flat whereas the COSMOS and JADES fields have (by design) sub-regions that differ significantly in depth. In COSMOS we found it useful to define 2 individual sub-regions corresponding to the deeper core and the wider/shallower outer region which we name ``COSMOS Deep'' and ``COSMOS Wide'' respectively. In JADES we define 3 distinct sub-regions. ``JADES Deep'' comprises the deeper stripes in the north of the imaging as well as the deep pointing in the South often labelled as the JADES Origins Field \citep[JOF;][]{robertson2023b}. ``JADES Medium'' and ``JADES Shallow'' define the remaining area. In Table \ref{tab:depths_jwst} we list the $5\sigma$ global depths for each field where COSMOS and JADES are divided into their respective sub-regions. The depths have been corrected to total assuming a point source correction. The areas of each region are also stated. 

To ensure consistent and robust high redshift galaxy selection, we restricted the analysis (i.e. the source selection) area  in each field to those regions in which deep \textit{HST}/ACS F814W and \textit{JWST}/NIRCam F090W imaging is available; this guarantees that there are always a sufficient number of "short"-wavelength filters to confirm the anticipated non-detections for $z\geq8.5$ galaxy candidates. In practice this requirement only has a significant effect on the NGDEEP and UDS fields, with the useful area of the {\it JWST} imaging in both these fields being reduced by approximately 1/3. This reduction in usable area has been accounted for in the areas noted in Table~\ref{tab:depths_jwst}.

\begin{table*}
        \centering
	\caption{The derived $5\sigma$ global depths for all the images used in this analysis. All depths (given in AB mag) have been measured in 0.3-arcsec diameter apertures on the PSF-homogenised images and then corrected to total assuming a point-source correction.}
	\label{tab:depths_jwst}
	\begin{tabular}{lccccccc} 
		\hline
         & PRIMER & PRIMER & PRIMER & & & \\
		Filter & COSMOS Deep & COSMOS Wide & UDS & NGDEEP & JADES Deep & JADES Medium & JADES Shallow\\
            \hline
            F435W & 28.0 & 27.8 & 27.2& 28.6 & 28.5 & 28.4 & 28.4\\
		F606W & 27.9 & 27.8 & 28.0& 29.3 & 28.9 & 28.9 & 28.9\\
        F775W & -& - & - & 28.9 & 28.2 & 28.1 & 28.2\\
		F814W & 28.0 & 27.6  & 29.0& 28.7 & 28.4 & 28.4 & 28.4\\
        F850LP & - & -  & - & 28.2 & 27.9 & 27.8 & 27.8\\
		\hline
		F090W & 28.0 & 27.6 & 27.6& - & 29.4 & 29.0 & 28.6\\
		F115W & 28.3 & 27.9 & 27.8& 29.1 & 29.8 & 29.4 & 29.0\\
		F150W & 28.4 & 28.1 & 28.1& 29.1 & 29.8 & 29.4 & 29.1\\  
		F200W & 28.6 & 28.2 & 28.2& 29.1 & 29.9 & 29.5 & 29.2\\
		F277W & 28.7 & 28.3 & 28.3& 29.5 & 30.2 & 29.8 & 29.3\\
		F356W & 28.8 & 28.4 & 28.3& 29.3 & 30.1 & 29.7 & 29.3\\
		F410M & 28.2 & 27.7 & 27.7& - & 29.7 & 29.1 & 28.7\\
		F444W & 28.4 & 28.0 & 27.9& 29.3 & 29.0 & 28.8 & 28.5\\
		\hline
        Area [sq. arcmin] & 47.8 & 84.4 & 170.7 & 5.6 & 19.6 & 25.3 & 15.8\\
        \hline
        
	\end{tabular}
\end{table*}

Due to the varying depths across much of the imaging, the global depth is a rather poor representation of the uncertainty in the flux of individual objects. To improve on this we have adopted the $1\sigma$ {\it local} depth as the uncertainty of the measured fluxes in each filter for each object. To establish the local depths we use the same technique as utilised by  \citet{donnan2023a,donnan2023b} in which the depth is calculated using the 200 empty apertures that lie closest to each detected source. 

\section{Sample selection}
\label{sec:sample}
For each field we constructed three catalogues using \textsc{SExtractor} \citep{bertin1996} in dual-image mode. We performed a rest-frame UV selection by using the F150W, F200W and F277W images as the detection images, as these filters sample the flux long-ward of the Lyman break at $9 \lesssim z \lesssim 20$. Master catalogues were constructed from the three single-filter catalogues, removing duplicates by retaining the duplicate with the highest signal-to-noise in its respective detection image. Photometry was measured within 0.3$^{\prime\prime}-$diameter circular apertures on the PSF homogenised images, corresponding to $\sim 70$ per cent of the total flux for a point source. Based on the individual curves of growth, further corrections of the order $\sim 1 - 2$ per cent were made to correct the imaging through every filter to exactly $70$ per cent of the total flux in order to improve the photometric homogenisation.

\subsection{SED fitting}
We used \textsc{eazy} \citep{brammer2008} to  perform our intial SED fitting, exploring the redshift range $0<z<20$ with the \textsc{Pegase} \citep{pegase1999} template set that includes nebular emission lines.

When computing the UV LF, the standard approach is to adopt the maximum likelihood (i.e. minimum $\chi^2$) 
photometric redshift solution for each source. However, this approach makes the implicit assumption that the integrated $p(z)$ within the adopted redshift bin is unity. This assumption can hold true when the signal-to-noise of the photometry is sufficient to exclude any alternative redshift solutions as being statistically unacceptable fits to the data. However, this is not the case for a significant number of the sources initially selected, particularly if they are close to the detection limit of the imaging. Consequently, a more robust approach to calculating the UV LF is to consider the full $p(z)$ of each source in the initial catalogue over the entire redshift range $0<z<20$. 

\subsection{Posterior probability distributions}
\label{sec:prior}
\begin{figure}
	\includegraphics[width=\columnwidth]{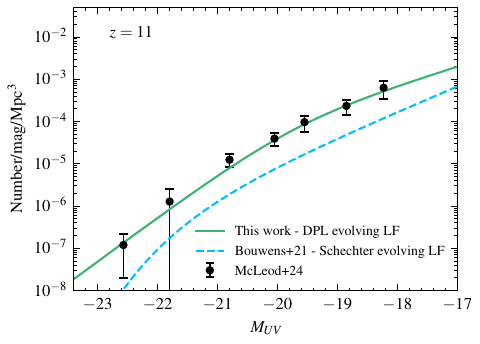}
    \caption{A comparison of our evolving double power-law parameterisation of the 
    $z = 11$ UV LF (green solid line) with the observational data from \citet{mcleod2023} at the same redshift. The prediction of the evolving Schechter function parameterisation of the UV LF from \citet{bouwens2021} is shown as the dashed blue line.}
    \label{fig:prior_LF}
\end{figure}

For each galaxy, we assume that the posterior probability distribution of the redshift, given the observed fluxes ($F$), is given by
\begin{equation}
    p(z|F) = \mathcal{L}(F|z) p(z|\rm{M_{UV}}),
\end{equation}
where $\mathcal{L}(F|z)$ is the likelihood of the observed fluxes given the redshift,
taken to be
\begin{equation}
    \mathcal{L}(F|z) \propto e^{-\chi^{2}(z)/2},   
\end{equation}
and $p(z|\rm{M_{UV}})$ is the prior probability of the redshift based on the implied absolute UV magnitude. This prior is based on an evolving model of the UV LF from $z=0-20$. At redshifts $z<7$ we adopt the evolving Schechter function parameterisation from \citet{bouwens2021}. 
At $z \geq 7$ we adopt a new parameterisation, designed to reproduce the double power-law (DPL) fits 
to the $z\simeq7-11$ UV LF from \citet{bowler2017,bowler2020,donnan2023a,mcleod2023}. The evolution of the DPL parameters is described as follows: 
\begin{gather}
    M^* = -20.95+0.11z\\    
    \phi^* = 10^{(-0.14z-2.36)}\\   
    \alpha = -2.04\times 10^{-4}z-2.1 \\   
    \beta = 0.138z-5.13.
\end{gather}
Fig.~\ref{fig:prior_LF} demonstrates that our model of the evolving UV LF is a good fit to the observational data at $z\simeq11$ from \citet{mcleod2023}. In contrast, it can be seen that an extrapolation of the Schechter model from \citet{bouwens2021} under predicts the latest \textit{JWST} observational data at these redshifts. 

\begin{figure}
	\includegraphics[width=\columnwidth]{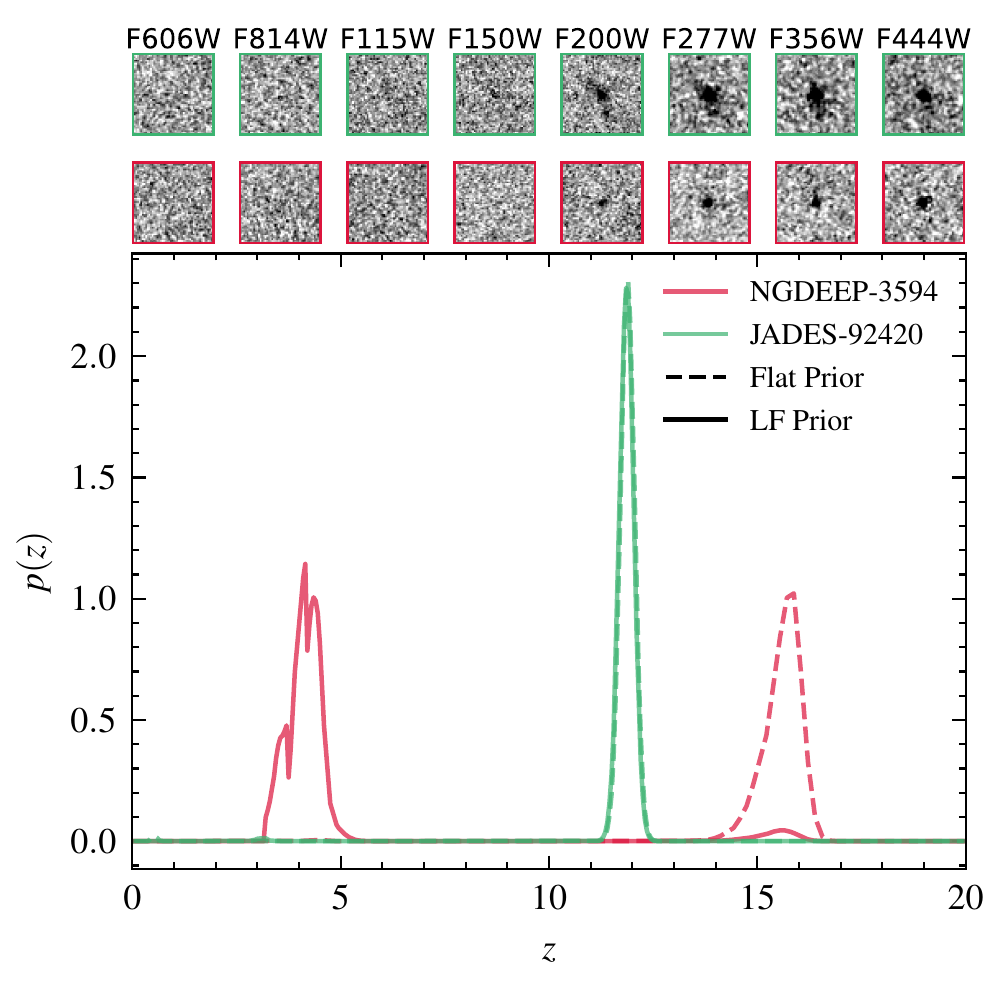}
    \caption{The posterior redshift probability distribution for NGDEEP-3594 (red) and JADES-92420 (green) using a flat redshift prior (dashed-lines) and our UV LF prior (solid-lines). The first source is a $z\sim16$ candidate reported by \citet{austin2023} and \citet{leung2023}, while the second is a spectroscopically-confirmed galaxy at $z=11.6$ \citep{curtislake2022}. The UV LF prior strongly weights the posterior redshift probability distribution to lower redshift for the less robust source, whereas the redshift solution of the robust high-redshift galaxy is unaffected. The top panel shows cut-out images of the two sources (indicated by their respective border colours) in the \textit{HST}/ACS F606W, F814W filters and the \textit{JWST}/NIRCam F115W, F150W, F200W, F277W, F356W and F444W filters. } 
\label{fig:priors}
\end{figure}

The impact of the UV LF prior is particularly important for very bright and/or extreme redshift sources, for which the posterior redshift probability distribution becomes weighted towards lower redshift {\it if the photometric data are not deep enough to robustly exclude such solutions}. In Fig.~\ref{fig:priors} we demonstrate the the effect of the UV LF prior on the posterior redshift distribution for two different sources: a galaxy uncovered in the HUDF which now has a spectroscopically-confirmed redshift of $z=11.6$, and an object with a photometric redshift of $z \sim 16$ selected from the NGDEEP survey. We also show their images in a number of filters.

The first source is a galaxy that was initially identified using \textit{HST} data in the HUDF \citep{bouwens2011,ellis2013,mclure2013} with only a single-band detection implying an extreme photometric redshift of $z \simeq 12$. This redshift solution has since been refined first with photometry from \textit{JWST}/NIRCam \citep{bouwens2023a,donnan2023b,robertson2022} and subsequently with spectroscopy from \textit{JWST}/NIRSpec \citep{curtislake2022} which yielded a robust redshift of $z=11.6$. Fig.~\ref{fig:priors} demonstrates that the $p(z)$ we derive for this source (JADES-92420) from fitting our {\it JWST} photometry is actually unaffected by the UV LF redshift prior due to the robust nature of its photometric redshift. This source is robustly at $z>8.5$ with an integrated $p(z\geq8.5)=0.995$. It most strongly contributes to the $z=12.5$ UV LF bin as described in Section~\ref{sec:LF} with $p(11.5<z<13.5)=0.97$.

The second source was identified by \citet{austin2023} to be at $z_{\rm{phot}}\simeq15.6$ (NGD-z15a) which, if correct, would make it one of the most distant galaxy candidates discovered to date. This source was also reported in \citet{leung2023}, with a very similar photometric redshift estimate of $z_{\rm{phot}}\simeq15.8$ (NGDEEP 1369). It can be seen from Fig.~\ref{fig:priors} that, with a flat redshift prior, we would also identify  this source (here listed as NGDEEP-3594) as a robust $z\geq 15$ candidate based on our photometry. However, the application of our adopted UV LF prior changes the posterior probability distribution in a way which strongly favours the lower-redshift solution at $z\sim4$,  indicating that, given the available data, this candidate is unlikely to be at ultra-high redshift, with a integrated $p(z\geq8.5)=0.054$. Although this candidate does still contribute to the tentative $z=14.5$ bin described in Section~\ref{sec:LF}, it only contributes the equivalent of 0.025 galaxies. This situation may appear similar to that of the very luminous $z\simeq16$ candidate detected in CEERS \citep{donnan2023a,finkelstein2022c} where subsequent spectroscopy with \textit{JWST}/NIRSpec conclusively revealed that the redshift was in fact $z=4.9$ \citep{arrabalharo2023} where the photometry was dominated by strong rest-frame optical emission lines \citep{harikane2023b}. However, the $p(z)$ for this highly-unusual candidate was in fact unaffected by the application of a UV LF prior, due to the relatively high signal-to-noise of the {\it JWST} photometry. Instead, the introduction of an extreme emission-line template SED was required to reveal the correct photometric redshift for this particular source.

\subsection{Selection of galaxy candidates}
\label{sec:selection}
Based on the master photometric catalogues for each field, we produced initial samples of galaxy candidates at $8.5<z_{\rm{phot}} \lesssim 20$. To be included in our initial samples, objects were required be $<2 \sigma$ detections in all ACS filters and the F090W NIRCam filter. Objects were also required to be a $\geq 5\sigma$ detection in one of the three detection filters: F150W, F200W and F277W as well as a $\geq 3 \sigma$ detection in any one of the other available NIRCam filters. To minimise contamination from artefacts, we masked the low SNR regions around the edges of each field, together with bright stars and their associated diffraction spikes. These further corrections/reductions are accounted for in the areas noted in Table~\ref{tab:depths_jwst}.

The next step in the selection process was to calculate the posterior redshift distribution for each source (see Eqn.~1), using the UV LF prior. The implementation of 
the UV LF prior required knowledge of the best-fitting value of $M_{\rm UV}$ at each redshift, which was calculated 
using a top-hat filter centred on 1500\,\AA\,in the rest-frame of the best-fitting SED. 
In order to calculate the total $M_{\rm{UV}}$, the aperture-based fluxes were scale to a Kron aperture flux \citep{kron1980}, with an additional correction of 10 per cent to
account for extended flux not accounted for by the Kron aperture \citep{mcleod2023}.

All objects with an integrated posterior redshift distribution of $p(z|F)\geq 0.05$ at $z\geq8.5$ were kept as viable high-redshift candidates. The final stage in the selection process was a visual inspection of the 
sample in order to remove the small number of remaining artefacts and diffraction spikes. 

\begin{figure}
	\includegraphics[width=\columnwidth]{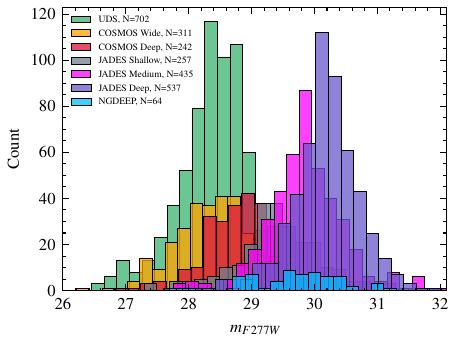}
    \caption{The distribution of apparent magnitude in the F277W filter for the galaxies in the final sample, split by the field in which they reside. The histogram is presented in bins with a width of 0.2 mag and the number of galaxies selected from each field is noted in the legend. The final total combined sample contains 2548 galaxies.}
    \label{fig:sample}
\end{figure}

The apparent magnitude distribution of the final sample in the F277W filter is shown in Fig.~\ref{fig:sample} across 
the six survey areas employed in this study.

\section{UV Luminosity function}
\label{sec:LF}
The galaxy selection process produced a final sample of 2548 galaxies selected over a total area of $\simeq 369$ square arcminutes of \textit{JWST}/NIRCam imaging. In this section we use this sample to determine the UV luminosity function at $8.5<z<15.5$.

\subsection{Completeness}
The accurate determination of the UV LF requires that the incompleteness in the final galaxy sample is accounted for. To achieve this we employed a simulation designed to mimic as closely as possible to real galaxy selection process. In order to calculate the fraction of recovered sources as a function of $M_{\rm UV}$ and $z$, artificial sources were injected into the real imaging for each of the fields over the range $-22<M_{\rm UV}<-17$ and $8.5<z<19.5$. This included separating COSMOS and JADES into their respective sub-fields as noted in Section~\ref{sec:depths}. 
In steps of $\Delta m=0.5$ and $\Delta z=0.5$, sources were injected into all of the available images for each field as point sources, based on an SED template with a UV-slope typical of the $z\geq8.5$ population \citep[$\beta \sim -2.2$;][]{cullen2023,morales2023,topping2023}. 

By performing the completeness simulation in this way, we are able to test every step of our selection process, ensuring that the recovered completeness is as accurate as possible. Sources were injected as point sources for simplicity, as detailed simulations have shown that modelling the physical size distribution of the sources has a negligible impact on the recovered completeness at these redshifts \citep{mcleod2023}. 
After injecting the sources, catalogues were extracted from the injected images in the same manner as for the main galaxy sample and then passed through the same selection process. The fractional completeness at each point on the $M_{\rm UV}-z$ plane was then calculated for each field. 

\subsection{Calculating the UV luminosity function}
In order to calculate the UV luminosity function, we populate the $M_{\rm{UV}}-z$ plane based on 
the normalized posterior redshift probability distribution of all galaxies in the final sample. In practice, we split the $M_{\rm{UV}}-z$ plane into bins of dimension $\Delta m=0.5$ and $\Delta z=0.5$ within the range $-22<M_{\rm UV}<-17$ and $8.5<z<19.5$, in order to match the resolution of the completeness simulation. The population of the $M_{\rm{UV}}-z$ plane was calculated for each sub-field individually, before the fields were combined to mimic a single survey. The total combined number density at each $M_{\rm UV}$ and $z$ is therefore given by,
\begin{equation}
    \Phi(M_{\rm UV},z) = \sum\limits^{N}_{i=1}\frac{p_i(M_{\rm UV},z)}{V_i C_i(M_{\rm UV},z)}
\end{equation}
where $N=7$ (the total number of sub-fields); $p_i(M_{\rm UV},z)$ is the total probability in a given $(M_{\rm UV}, z)$ bin for a given field; $C_i(M_{\rm UV},z)$ is the corresponding completeness and $V_i$ is the cosmological volume provided by that field. When computing the combined UV LF, we conservatively restrict the contribution of each sub-field to the UV magnitude range where it is $\geq50$ per cent complete. The final result provides a continuous expression of the UV LF in two dimensions, from which the one-dimensional UV LF can be extracted over any chosen range in redshift. As a result, and unlike many literature studies, it is not necessary to calculate the
the UV LF over wide redshift bins in order to counteract the impact of photometric-redshift uncertainties.

\begin{figure*}
	\includegraphics[width=\textwidth]{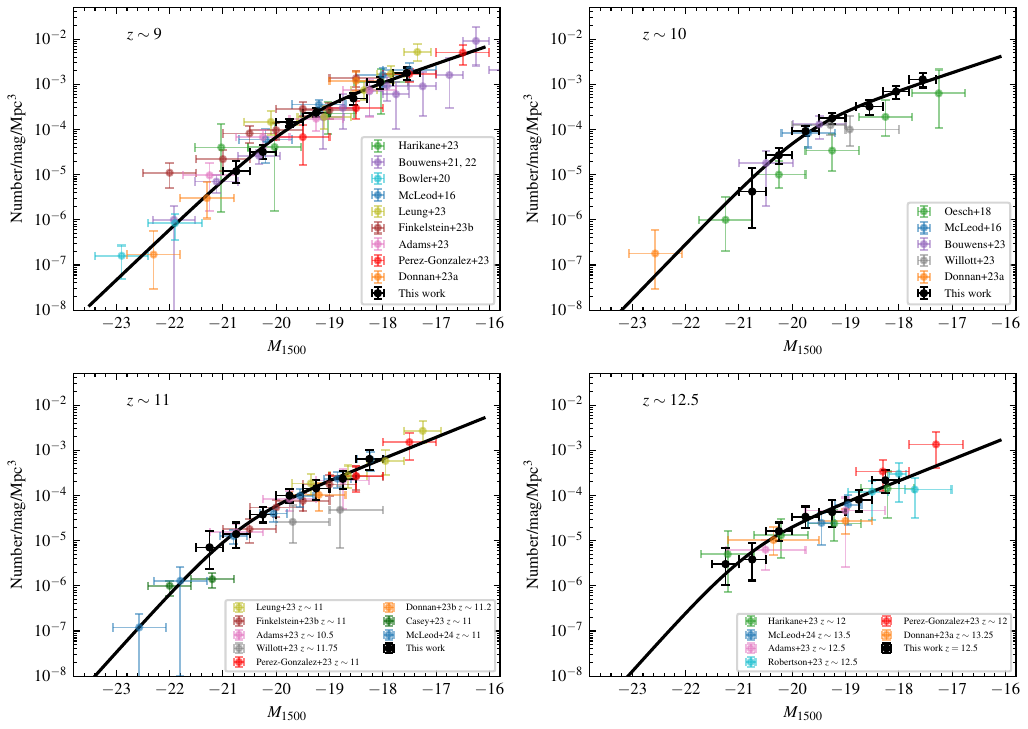}
    \caption{Our new measurements of the rest-frame UV LF at $z=9, 10, 11$ and $z=12.5$ are shown as the black data points. For comparison we also show data points from \citet{mcleod2016,oesch2018,bowler2020,bouwens2021,bouwens2022,harikane2023a,adams2023,leung2023,donnan2023a,donnan2023b,mcleod2023,perezgonzalez2023,finkelstein2023,casey2023,bouwens2023b, willott2023, robertson2023b} with colours indicated in the figure legend. The best fitting double-power law functions are shown as the solid black lines at each respective redshift.}
    \label{fig:LF}
\end{figure*}

We extract and plot (in Fig.~\ref{fig:LF}) the one-dimensional UV LF centered on $z=9, 10, 11$ and $z=12.5$ using redshift bins spanning $8.5<z<9.5$, $9.5<z<10.5$, $10.5<z<11.5$ and $11.5<z<13.5$. The galaxy number densities are tabulated in Table~\ref{tab:LF_points} along with their corresponding uncertainties. The uncertainties were calculated using Poisson confidence intervals from \citet{gehrels1986} combined in quadrature with the cosmic variance uncertainty. The cosmic variance was estimated using the calculator from \citet{trenti2008} for each of the survey fields using the default cosmological parameters with a $\sigma_8=0.9$ and a halo-filling factor of unity. This was then combined using the prescription from \citep[equation (9) in ][]{moster2011}. However, the cosmic variance uncertainty (which ranges from $15\%$ to $25\%$) in fact has minimal impact on the final uncertainties and including them does not significantly alter our results. For comparison, in Fig.~\ref{fig:LF} we also plot a number of other measurements of the UV LF from the recent literature.

At $z=9$ our new measurement of the UV LF is consistent with the early {\it JWST}-based results of \citet{donnan2023a}, as well as with other recent \textit{JWST} and pre-\textit{JWST} studies \citep[e.g.][]{mcleod2016,bouwens2021}. We also find good consistency with the faint-end slope measurements from \citet{bouwens2022} which were based on an analysis of the available lensing fields from \textit{HST} which enabled a measurement of the UV LF down to $M_{\rm UV}\simeq-16$ at $z=9$. At $z=10$ there is good consistency between our new measurement and the pre-\textit{JWST} measurement from \citet{mcleod2016}, as well as an early analysis of \textit{JWST} imaging from \citet{bouwens2023b}. At $z=11$ we can compare to a significant number of recent \textit{JWST} studies and find that our new measurement of the UV LF is consistent with many of them, in particular the recent wide-area study of \citet{mcleod2023}. Our determination of the UV LF at $z=11$ is also consistent with measurements of the faint-end of the $z=11$ UV LF by \citet{leung2023} and \citet{perezgonzalez2023}, who studied a relatively small area ($\simeq 8$ sq. arcmin) covered by the NGDEEP and MIDIS fields, respectively. However, it can be seen that our new measurement of the $z=11$ UV LF is consistently higher than the recent determination by \citet{willott2023}.  

At $z=9, 10$ and $z=11$ we are able to achieve a dynamic range of $\simeq4$ magnitudes ($-21<M_{\rm UV}<-17$) due to our multi-field approach, without relying on gravitational lensing. 
This is transformative compared to the early measurements of the UV LF from \textit{JWST} and, in combination with our large sample size, allows the UV LF to be measured with significantly lower uncertainties. Overall we find a particular lack of evolution at $z=9-11$ which is consistent with previous \textit{JWST} studies \citep[e.g.][]{finkelstein2023}. At $z=12.5$ we are still able to measure the UV LF over a dynamic range of $\simeq 3$ magnitudes ($-21<M_{\rm UV}<-18$), significantly increasing the number of LF bins at this redshift. Our new determination of the $z=12.5$ UV LF is in reasonable agreement with recent \textit{JWST}-based measurements from the literature \citep[e.g.][]{adams2023,robertson2023b}.

\begin{figure}
	\includegraphics[width=\columnwidth]{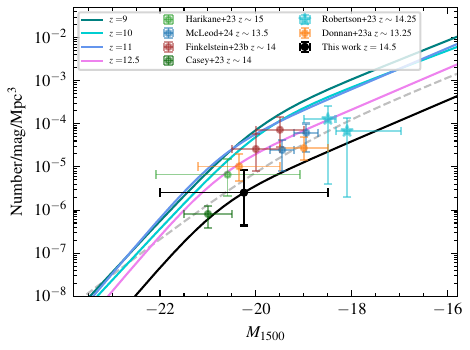}
    \caption{Our tentative measurement of the rest-frame UV LF at $z=14.5$ is shown as the black data point. We include data points from \citet{harikane2023a,donnan2023a,mcleod2023,finkelstein2023,casey2023,robertson2023b} for comparison. The best-fitting double power-law function is shown as the solid black line, where only the normalisation was allowed to evolve from $z=12.5$. The dashed grey line shows the prediction from the evolving UV LF model described in Section~\ref{sec:prior}. The solid coloured lines show our best fitting double power-law fits at $z=9,10,11$ and $z=12.5$, with colours as indicated in the figure legend.}
    \label{fig:LF_z14}
\end{figure}

With this sample we are able to robustly measure the evolution of the UV LF from $z=9-13$. However, beyond $z=13.5$ there is a noticeable dearth in the total $p(z)$ and therefore we are unable to precisely measure the UV LF at this redshift. That said, although limited by small-number statistics (the equivalent of $\simeq1.3$ galaxies), we are able to plot a single bin at $z\simeq14.5$, spanning the 
range $13.5<z<15.5$. This single bin is shown in Fig.~\ref{fig:LF_z14}, where we compare to measurements from other recent studies centred on redshifts from $z\simeq13-15$.

\begin{table}
	\centering
	\caption{Computed UV LF data points shown in Fig.~\ref{fig:LF} at $z=9, 10, 11, 12.5$ and in Fig.~\ref{fig:LF_z14} at $z=14.5$. The columns show the central redshift and central UV absolute magnitude of each bin, and then the source number densities within each bin, along with their corresponding uncertainties.}
	\label{tab:LF_points}
	\def\arraystretch{1.35}
	\begin{tabular}{lcc} 
		\hline
		$z$ & $M_{\rm UV}$ / AB mag  & $\phi$ / ${\rm 10^{-6}/mag/Mpc^{-3}}$\\
  \hline
9 & $-20.75$ & $\phantom{00}12^{+8\phantom{00}}_{-5\phantom{00}}$\\ 
9 & $-20.25$ & $\phantom{00}32^{+13\phantom{0}}_{-10\phantom{0}}$\\ 
9 & $-19.75$ & $\phantom{0}144^{+30\phantom{0}}_{-28\phantom{0}}$\\ 
9 & $-19.25$ & $\phantom{0}235^{+60\phantom{0}}_{-49\phantom{0}}$\\ 
9 & $-18.55$ & $\phantom{0}486^{+157}_{-139}$\\ 
9 & $-18.05$ & $1110^{+310}_{-310}$\\ 
9 & $-17.55$ & $1776^{+578}_{-510}$\\ 
\hline
10 & $-20.75$ & $\phantom{000}4^{+10\phantom{0}}_{-4\phantom{00}}$\\ 
10 & $-20.25$ & $\phantom{00}27^{+13\phantom{0}}_{-10\phantom{0}}$\\ 
10 & $-19.75$ & $\phantom{00}92^{+25\phantom{0}}_{-20\phantom{0}}$\\ 
10 & $-19.25$ & $\phantom{0}177^{+53\phantom{0}}_{-45\phantom{0}}$\\ 
10 & $-18.55$ & $\phantom{0}321^{+127}_{-111}$\\ 
10 & $-18.05$ & $\phantom{0}686^{+245}_{-223}$\\ 
10 & $-17.55$ & $1278^{+486}_{-432}$\\ 
\hline
11 & $-21.25$ & $\phantom{00}7^{+9\phantom{00}}_{-5\phantom{00}}$\\ 
11 & $-20.75$ & $\phantom{0}14^{+11\phantom{0}}_{-7\phantom{00}}$\\ 
11 & $-20.25$ & $\phantom{0}38^{+16\phantom{0}}_{-13\phantom{0}}$\\ 
11 & $-19.75$ & $100^{+37\phantom{0}}_{-30\phantom{0}}$\\ 
11 & $-19.25$ & $144^{+81\phantom{0}}_{-63\phantom{0}}$\\ 
11 & $-18.75$ & $234^{+118}_{-96\phantom{0}}$\\ 
11 & $-18.25$ & $641^{+361}_{-281}$\\ 
\hline
12.5 & $-21.25$ & $\phantom{00}3^{+4\phantom{00}}_{-2\phantom{00}}$\\ 
12.5 & $-20.75$ & $\phantom{00}4^{+5\phantom{00}}_{-3\phantom{00}}$\\ 
12.5 & $-20.25$ & $\phantom{0}16^{+9\phantom{00}}_{-6\phantom{00}}$\\ 
12.5 & $-19.75$ & $\phantom{0}34^{+23\phantom{0}}_{-15\phantom{0}}$\\ 
12.5 & $-19.25$ & $\phantom{0}43^{+35\phantom{0}}_{-22\phantom{0}}$\\ 
12.5 & $-18.75$ & $\phantom{0}80^{+51\phantom{0}}_{-36\phantom{0}}$\\ 
12.5 & $-18.25$ & $217^{+153}_{-104}$\\ 
\hline
14.5 & $-20.25$ & $\phantom{00}3^{+6\phantom{00}}_{-2\phantom{00}}$\\
\hline
	\end{tabular}
\end{table}

\subsubsection{Luminosity function fitting}
It is now typical to fit the UV LF at $z\geq7$ with a double-power law (DPL) rather than a Schechter function, due to the excess of very bright galaxies that have been detected from large area ground-based surveys \citep[e.g.][]{bowler2017,bowler2020,donnan2023a,varadaraj2023}. We therefore fit our new observational data points at $z=9, 10, 11$ and $z=12.5$ with a DPL parameterisation. We perform the fitting with the \textsc{Scipy} curve\_fit function \citep{virtanen2020} using a least-squares method to fit the data. To enhance the dynamic 
range in UV luminosity, we include in the fits the bright-end data points from \citet{bowler2020} and \citet{donnan2023a} at $z=9$ and from \citet{mcleod2023} at $z=11$. 

Given the limited constraints on the bright-end of the LF at $z=10$, we fix the bright-end slope to $\beta=-4.05$, as this is the midpoint between the best-fitting values at $z=9$ and $z=11$. Due to the more limited number of data points available at $z=12.5$, there is insufficient 
dynamic range in UV luminosity to perform a free-fit to the data. Consequently, we fix the faint-end and bright-end slopes to their best-fitting values at $z=11$, but keep the LF normalisation ($\phi^*$) and the characteristic magnitude ($M^*$) as free parameters. 

The results of the LF fits are shown as the solid black lines in Fig.~\ref{fig:LF} and the best-fitting DPL parameters are reported in Table~\ref{tab:dpl_params}. We also fit a DPL to the single bin at $z=14.5$, fixing $M^*$, $\alpha$ and $\beta$ to their best-fitting values at $z=12.5$, and fitting $\phi^*$ alone. This fit is shown as the black solid line in Fig.~\ref{fig:LF_z14}.

\begin{table*}
	\centering
	\caption{The derived parameter values for the best-fitting double power-law (DPL) models fitted to our data over the redshift range $9 < z < 15.5$. The LF fits derived at $z=9$ and $z=11$ utilised the new data presented here along with the data-points presented by \citet{bowler2020} and \citet{mcleod2023}. At $z=10$ and $z=12.5$ the fits are based purely on the new analysis and galaxy samples presented in this work. The first column gives the central redshift of the binned LF. This is followed by the values of the best-fitting characteristic density $\phi^*$, the best-fitting or fixed characteristic absolute magnitude $M^*$, the fitted or assumed faint-end slope $\alpha$, and the fitted or adopted bright-end slope $\beta$ (see text for details). In the cases where a parameter was fixed, the value is denoted with an asterisk. The final column states the resulting UV luminosity density derived at each redshift. }
	\label{tab:dpl_params}
    \setlength{\tabcolsep}{4pt} 
	\renewcommand{\arraystretch}{1.15} 
	\begin{tabular}{lccccc} 
		\hline
		$z$ & $\phi^* {\rm /mag/Mpc^{-3}}$ & $M^*$ /AB mag & $\alpha$ & $\beta$ & $\log_{10}(\rho_{\rm{UV}}/ \rm{ergs\, s^{-1} Hz^{-1} Mpc^{-3}})$\\
		\hline
		9 & ($23.5\pm39.2$)$\times 10^{-5}$ & $-19.70\pm0.96$ & $-2.00\pm0.47$ & $-3.81\pm0.49$ & $25.29^{+0.05}_{-0.05}$\\
        10 & ($14.5\pm16.4$)$\times 10^{-5}$ & $-19.98\pm0.61$ & $-1.98\pm0.40$ & $-4.05^*$ & $25.12^{+0.07}_{-0.08}$\\
		11 & ($3.27\pm9.86$)$\times 10^{-5}$ & $-20.73\pm1.61$ & $-2.19\pm0.69$ & $-4.29\pm$1.30 & $25.12^{+0.14}_{-0.20}$\\
		12.5 & ($0.99\pm0.99$)$\times 10^{-5}$ & $-20.82\pm0.71$ & $-2.19^*$ & $-4.29^*$ & $24.64^{+0.18}_{-0.32}$\\
        14.5 & ($0.18\pm0.28$)$\times 10^{-5}$ & $-20.82^*$ & $-2.19^*$ & $-4.29^*$ & $23.92^{+0.27}_{-0.81}$\\

		\hline
	\end{tabular}
\end{table*}

\begin{figure}
	\includegraphics[width=\columnwidth]{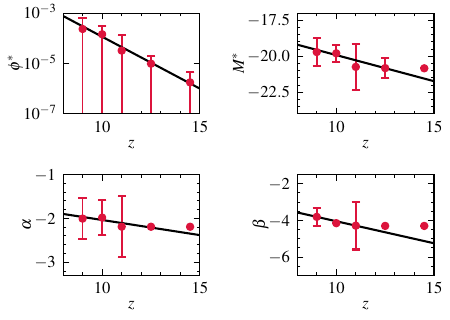}
    \caption{The evolution of the best-fitting double-power law parameters at $z=9,10,11,12.5$ from our UV LF fitting. The solid black line shows a linear fit to the evolving parameters. Where a parameter was fixed, its value is indicated by a data point with no error bar. }
    \label{fig:LF_params}
\end{figure}

\subsection{The cosmic star-formation rate density}
\begin{figure*}
	\includegraphics[width=\textwidth]{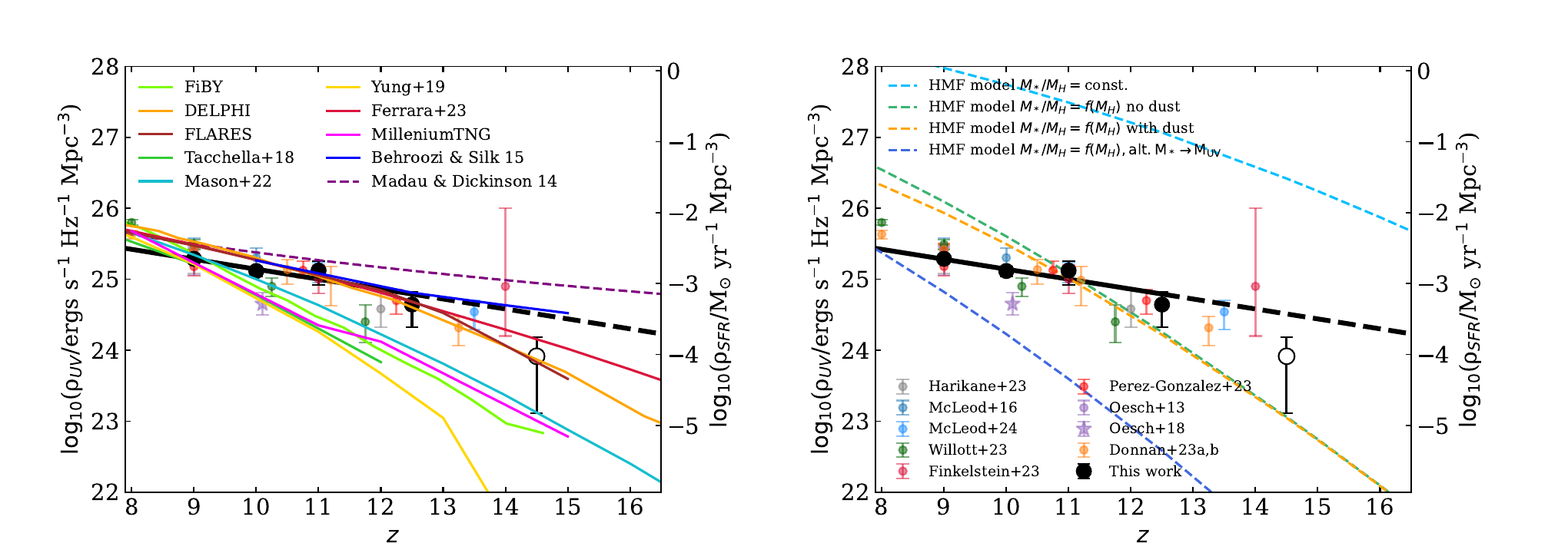}
    \caption{The redshift evolution of the UV luminosity density, $\rho_{\rm UV}$, and hence the inferred cosmic star-formation rate density, $\rho_{\rm SFR}$, at $z>8$ with our new measurements at $z\sim9,10,11,12.5$ (solid black circular data points) and our tentative measurement at $z=14.5$ (open black circular data point). Estimates at $z\simeq9-10$ from \citet{oesch2013,oesch2018} and \citet{mcleod2016} are shown by the purple and blue data points respectively. The green, grey and light red points show $\rho_{\rm UV}$ derived from the LFs in \citet{willott2023}, \citet{harikane2023a} and \citet{perezgonzalez2023} respectively. All values were determined using a limit of $M_{\rm UV}=-17$ in the luminosity-weighted integral. The orange data points show the results from \citet[][]{donnan2023a,donnan2023b}. The solid black line shows the best-fitting linear relation at $z=9-12.5$ with an extrapolation shown by the dashed black line. The left-hand panel shows a comparison to theoretical models, again setting the LF integration limit to $M_{\rm UV}=-17$. The green, cyan, blue and yellow lines show semi-analytic models from \citet{tacchella2018}, \citet{mason2022}, \citet{behroozi2015} and \citet{yung2019} respectively. The red line shows the dust-free model from \citet{ferrara2023}. The green, brown and pink lines show the results of hydrodynamical simulations from FiBY \citep{johnson2013,paardekooper2015}, FLARES \citep{wilkins2022}, DELPHI \citep{mauerhofer2023} and MilleniumTNG \citep{kannan2022}. The dashed purple line shows an extrapolation at $z\geq8$ of the $\rho_{\rm{UV}} \propto (1+z)^{-2.9}$ relation from \citet{madau2014}. The dashed lines in the right panel show the predictions from our halo mass function (HMF) models. We show the HMF model where $\log(M_*/\rm M_{\odot})=9$ corresponds to $M_{\rm UV}=-22.47$ assuming a constant stellar mass to halo mass ratio (light blue), the same mass-to-light conversion but with a halo-mass dependent stellar mass to halo mass ratio consistent with that at $z=0$ (green) and then where dust is introduced (orange). The dark blue curve shows the same as the model in green but with an alternative scaling where $\log(M_*/\rm M_{\odot})=10$ corresponds to $M_{\rm UV}=-22.4$.} 
    \label{fig:rho_uv}
\end{figure*}

The UV luminosity density ($\rho_{\rm UV}$) was calculated by performing a luminosity-weighted integral of our best DPL fits to the data. We integrated down to a limit of $M_{\rm UV}=-17$ which is consistent with many other \textit{JWST} studies \citep[e.g.][]{finkelstein2023,harikane2023a,adams2023}. The UV luminosity density was then converted to the cosmic star-formation rate density ($\rho_{\rm SFR}$) using the conversion factor $\cal{K}$$_{\rm UV} = 1.15 \times 10^{-28}$ M$_{\odot}$ yr$^{-1}$/erg s$^{-1}$ Hz$^{-1}$ \citep{madau2014} which assumes a \citet{salpeter1955} initial mass function (IMF). The results are shown in Fig.~\ref{fig:rho_uv}. 

The improved statistics and dynamic range in UV luminosity of this study provide measurements 
of the UV luminosity density at $z=9$ to $z=12.5$ with much lower uncertainties than early \textit{JWST-}based measurements. 

In \citet{donnan2023a} we performed a log-linear fit to the evolution of $\rho_{\rm UV}$ with redshift, motivated in part by the analytical expression from \citet{hernquist2003}. An updated log-linear fit to our new data is shown as the solid black line in Fig.~\ref{fig:rho_uv} and has the functional form
\begin{equation}
    \log(\rho_{\rm UV}) = (-0.140 \pm 0.068) z + (26.5 \pm 0.6).
\end{equation}
It can be seen that this expression provides an excellent description of the evolution of $\rho_{\rm UV}$ over the redshift range $8<z<12.5$. Interestingly, it is noticeable that an extrapolation of this relation (dashed black line) sits somewhat higher than our tentative measurement of $\rho_{\rm UV}$
at $z=14.5$.

\section{Discussion}
\label{sec:discussion}
\subsection{A multi-field approach to measuring the galaxy LF}
By combining the data from the PRIMER, JADES and NGDEEP {\it JWST} NIRCam surveys we have been able to conduct a relatively wide-area survey which also samples $\simeq4$ mag of dynamic range in UV luminosity at $z=9$ to $z=12.5$.  This, coupled with a rigorous, statistically robust methodology, has enabled us to substantially improve our knowledge of the form and evolution of the UV galaxy LF at $z\geq9$ compared to what was achieved by previous studies. 

Firstly, by combining four separate survey fields with a total combined area of $\simeq370$ sq. arcmin, we have been able to mitigate the impact of cosmic variance on our measurements. Primarily due to the impact of PRIMER, our survey area is an order-of-magnitude larger than that which was available in the initial studies of the high-redshift UV LF undertaken with \textit{JWST} data. For example, the first measurements of the UV LF with \textit{JWST} were derived  from early imaging which covered only $\sim 40$ sq. arcmin  \citep{donnan2023a,finkelstein2022b,bouwens2023b}. As {\it JWST} Cycle 1 progressed, the area of available imaging expanded somewhat, to $\sim 100$ sq. arcmin \citep{harikane2023a,adams2023,finkelstein2023}. However, all of these early NIRCam-based studies essentially utilised the same survey fields, notably the CEERS \citep{bagley2023} ERS imaging of the EGS field (initially covering $\simeq35$ sq. arcmin, subsequently expanding to $\simeq90$ sq. arcmin after programme completion) and the Abell 2744 cluster field which was imaged by the GLASS \citep{treu2022} and UNCOVER \citep{bezanson2022} programmes. As noted in \citet{castellano2023}, as well as being relatively small (and complicated by gravitational lensing), this latter field has a significant over-density at $z\simeq10$, severely exacerbating the impact of cosmic variance on LF determinations at these redshifts. 

These limitations  motivated the work of \citet{mcleod2023}, which represents the most extensive study of the UV LF at $z=11$ completed prior to the present work. This study was based on a combined survey area of $\simeq210$ sq. arcmin, and the results solidified the earlier measurements of the extreme-redshift UV LF, with the data again supporting a slow, gradual evolution of the LF between $z=9$ and $z=11$. The results presented here are in excellent agreement with those derived by  \citet{mcleod2023},  which is all the more significant because the present study is based on a different set of survey fields, and utilises a different method for calculating the evolving LF. These consistent results, now based on effectively  $\simeq600$ sq. arcmin of NIRCam imaging, contrast with the evolution of the LF recently presented by \citet{willott2023}. This study used imaging from the CANUCS survey targeting five cluster fields, covering a relatively small total area of $\simeq 50$ sq. arcmin, with the medium-band imaging (a strength of the CANUCS dataset) only covering $\simeq 35$ sq. arcmin. Therefore, as they note, they may have targeted one or more relatively underdense fields, which most likely explains why their results differ somewhat from the higher number densities found here and in other wider area studies \citep[e.g.][]{mcleod2023,finkelstein2023}.

Secondly, our new study benefits from vastly improved statistics, simply because our combined multi-tiered survey has yielded a very large sample of galaxies from which to compute the UV LF over a wide range in UV luminosity and redshift. This has led to a significant reduction in the statistical  uncertainties in our measurements as well as enabling us to produce LF measurements in an increased number of redshift bins over the redshift range $z=9-12.5$. By increasing the dynamic range in UV luminosity compared to previous \textit{JWST} studies, we are better able to constrain the shape of the LF and hence explore how the parameters of the adopted DPL function evolve with redshift. In performing the DPL fits, we found that the data were of sufficient quality to enable us to allow the characteristic magnitude, $M^*$, and the characteristic density, $\phi^*$, to be fitted as free parameters at $z=9-12.5$. The results as a function of redshift are shown in the upper two panels of Fig.~\ref{fig:LF_params}, where it can be seen that the derived evolution in $\phi^*$ is stronger than that inferred for $M^*$.  This indicates that there is more density evolution than luminosity evolution in the UV LF over the redshift range $z=9-12.5$, consistent with the persistence of relatively bright galaxies out to extreme redshifts \citep[e.g.][]{castellano2023,mcleod2023}. 

We can also explore the evolution of the faint- and bright-end slopes of the DPL fit. Although the bright-end slope, $\beta$, is (by necessity) fixed for the fits at $z=10$ and $z=12.5$, we see a lack of evolution between the free fits achieved at $z=9$ and $z=11$. The same is also true of the faint-end slope, $\alpha$, where we find no significant evolution between $z=9$ (where we obtain $\alpha=-2.00\pm0.47$) and $z=11$ (where we find $\alpha=-2.19\pm0.69$). Our results are consistent with other recent studies of the faint-end slope, which also observe no significant change over this redshift range \citep{leung2023,perezgonzalez2023}. 

Using wide-area ground-based surveys, such as UltraVISTA, it has been demonstrated that there is little if any evolution in the bright end of the UV LF from $z\simeq7$ to $z \simeq 10$ \citep[e.g.][]{stefanon2019,bowler2020,donnan2023a}. However, due to the wavelength restriction of ground-based telescopes, the bright-end of the UV LF at $z>10$ can only be measured with wide-area surveys from \textit{JWST}. Our results re-affirm the findings of \citet{mcleod2023} and now extend the evidence for the lack of evolution in the bright-end of the LF out to to $z\simeq12.5$. Several previous studies have discussed the potential  physical mechanisms that might allow/explain this (arguably unexpected)  lack of evolution in the bright galaxy population at early times \citep[e.g.][]{bowler2017,bowler2020,finkelsteinBag2022}. A decrease in dust attenuation or a lack of AGN feedback have been suggested as contributing factors, while others have inferred that the data imply increased star-formation efficiency  in the very young Universe \citep{harikane2023a}. However, as we discuss further below, while these proposed astrophysical changes may be important, and can certainly not be excluded at present, they are not in fact required to explain the results of the present study.

In the left-hand panel of Fig.~\ref{fig:rho_uv} we compare our new measurements of $\rho_{\rm SFR}$ to a number of theoretical models and cosmological simulations. It can be seen that the model predictions diverge widely beyond $z \simeq 9$, and so the diagnostic power of the new measurements is clear. In particular, the observational data now clearly lie above the constant star-formation efficiency models of \citet{tacchella2018}, \citet{yung2019} and \citet{mason2022}. However, there is good agreement between the observations and the predictions of the FLARES \citep{vijayan2021,lovell2021,wilkins2022} and DELPHI \citep{mauerhofer2023} cosmological hydrodynamical simulations. Our results are also consistent with the predictions of the semi-empirical dust-free model presented by \citet{ferrara2023}, which is discussed further in Section~\ref{sec:modelling}. It should be noted that the DELPHI and \citet{ferrara2023} dust-free models were both published after initial \textit{JWST} studies suggested a high abundance of luminous galaxies at $z\geq10$.

\subsection{The star-formation rate density at $\mathbf{z\geq13}$}
The wealth of new deep NIRCam imaging delivered by  \textit{JWST} has already enabled a large number of galaxy candidates to be detected up to $z\simeq12$, and several of these extreme redshift candidates have already been spectroscopically confirmed with NIRSpec \citep[e.g.][]{harikane2023b,arrabalharo2023}. Indeed, as shown in the present work, the galaxy samples that can now be assembled at $z \simeq 12$ (by combining the major Cycle-1 programmes) are large and robust enough to enable the basic form and amplitude of the UV LF to be well constrained at these redshifts. However, the nature of galaxy evolution at still higher redshifts, $z\geq13$, remains much more uncertain. Despite the abundance of \textit{JWST} imaging now available, and the absence of any wavelength limitation, there remain very few robust galaxy candidates at $z\geq13$, with only one spectroscopic confirmation of a faint galaxy at $z=13.2$ \citep{curtislake2022}. This has inevitably led to uncertainty in constraining the UV galaxy LF and hence cosmic star-formation rate density at $z\geq13$.

 \citet{donnan2023a} proposed a log-linear relationship between comoving cosmic star-formation rate density, $\rho_{\rm{SFR}}$, and redshift, $z$, at $z\geq8$, motivated at least in part by the theoretical expectations articulated by \citet{hernquist2003}. Indeed, the early {\it JWST} results did appear consistent with the inferred smooth, gradual decline in $\rho_{\rm{SFR}}$ to higher redshift, and the results of the present study re-affirm this conclusion, showing that the log-linear relation remains a good fit to our new robust constraints over the redshift range  $z=9-12.5$, as shown by the solid black line in Fig.~\ref{fig:rho_uv}. However, at some point a departure from this relationship is inevitable as we enter the epoch of the very first galaxies. To explore to yet higher redshifts, 
as described in Section~\ref{sec:LF}, we have attempted to calculate a basic estimate of the galaxy number density at $z=14.5$ from our data. This measurement provides some tentative evidence for the onset of a steeper decline in the UV LF between $z=12.5$ and $z=14.5$, with the inferred $\rho_{\rm{SFR}}$ at $z=14.5$ lying below the extrapolation of the log-linear relation fitted at $z=9-12.5$. However, this measurement is highly uncertain, as indicated by the error bars, and we caution against any strong interpretation of this result given that our UV LF measurement at $z=14.5$ is based on a total $p(z)$ equivalent to $\simeq 1.3$ galaxies.

\citet{robertson2023b} also recently explored this very early epoch within the $\simeq8$ sq. arcmin of the JADES Origins Field (JOF). They report 3 candidates at $z>13.5$ but note that none of them can be regarded as robust. Therefore they discuss two alternative scenarios in which they calculate $\rho_{\rm{SFR}}$ with and without the 2 highest-redshift candidates (proposed to lie at $z>14$), yielding the expected result that there is a more rapid decline in $\rho_{\rm{SFR}}$ with the extreme redshift candidates removed. Although, as noted above, our own measurement of $\rho_{\rm{SFR}}$ at $z=14.5$ is somewhat poorly constrained, it is nonetheless more closely aligned with this more rapid decline scenario at $z>13$. Indeed, consistent with this, our own investigation of the JADES data supports the removal of all 3 of the $z>13.5$ candidates tentatively reported by \citet{robertson2023b}. We recover one of these galaxies in our sample (JADES+53.02868$-$27.89301) but find a $p(z)$ which peaks at $z\sim3.5$ (albeit still with a non-neglible probability of lying at $z\sim13$). The 2 candidates at $z>14$ tentatively reported by \citet{robertson2023b} do not contribute significantly to our high-redshift LFs because in both cases the bulk of their redshift probability distribution lies at much lower redshifts in our analysis (which we emphasize, however, involves the use of LF priors). This therefore adds to the tentative but growing evidence that there is a change from the gradual evolution in the LF observed at $z=9-13$ to a more rapid decline at $z>13$. However, still better constraints on the UV LF (and hence $\rho_{\rm{SFR}}$) are needed at these extreme redshifts to confirm the existence and/or severity of this transition.

\subsection{Modelling the growth of the galaxy population at $\mathbf{z>8}$}
\label{sec:modelling}
As discussed above, the initial observations from \textit{JWST} revealed a high abundance of (UV) bright galaxies at $z\geq10$ which stimulated  a number of theoretical attempts to explain their abundance. One natural point of astrophysical interest is the way in which the dust attenuation of galaxies might change ({\it i.e.} reduce) with increasing redshift. Indeed, \citet{ferrara2023} have recently proposed a model in which galaxies are essentially dust-free at very early times, finding good agreement with the first measurements of the UV LF from \textit{JWST}. In this specific model they propose that the dust could have been ejected from the galaxies as a by-product of intense star-formation activity, leading to (temporarily) dust-free galaxies that might populate the bright end of the LF at $z>10$. This model does indeed provide  a good fit to our new observational constraints on $\rho_{\rm{UV}}$, as shown by the red solid line in the left-hand panel of Fig.~\ref{fig:rho_uv}. There is now also independent observational evidence for a lack of significant dust in galaxies at $z>10$ as inferred from analyses of the UV continuum slopes, $\beta$, displayed by early galaxies. In particular, \citet{cullen2023} report that, at $z=11.5$, the average UV slope plateaus at  $\beta\simeq-2.6$, consistent with dust-free stellar populations. This is also consistent with the UV-slope measurements of \citet{morales2023} and \citet{topping2023}. It has also been suggested that increasingly stochastic star formation, with spells of enhanced star-formation efficiency, can help to explain the high number densities of galaxies observed in the rest-frame UV at $z\geq10$ \citep{mason2022}.

\begin{figure*}
	\includegraphics[width=\textwidth]{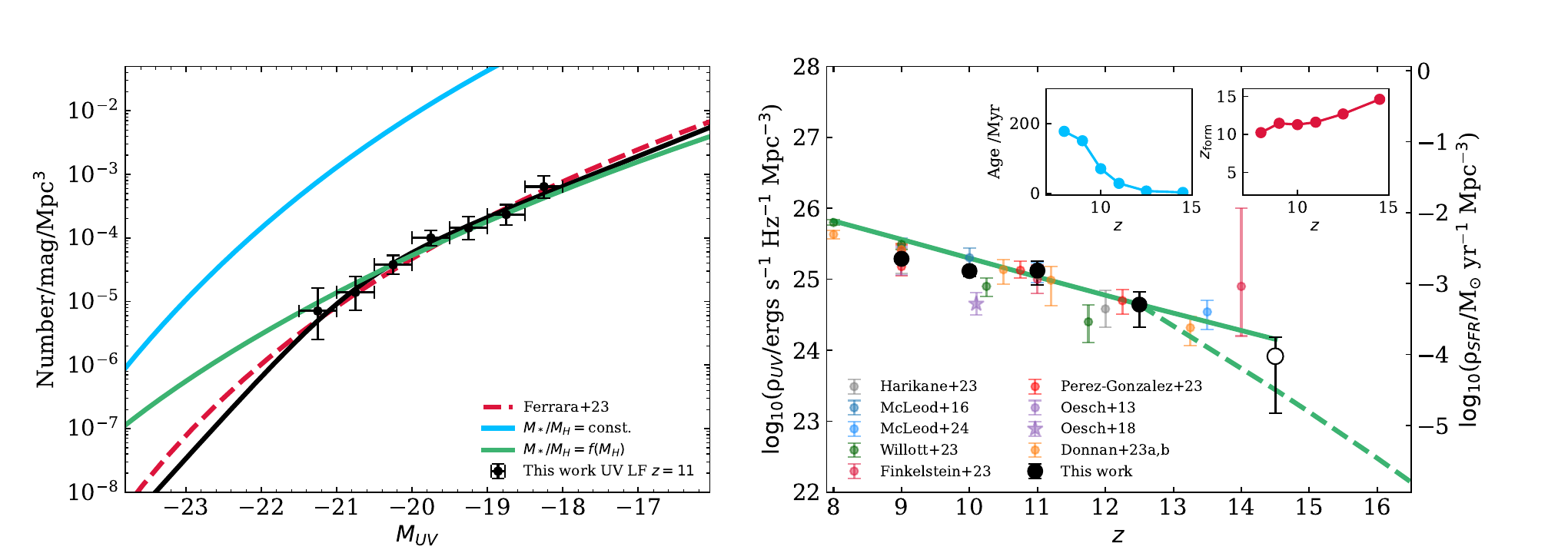}
    \caption{\textit{Left:} The observed UV LF at $z=11$ compared to our DPL fit (black), our HMF model assuming a constant $M_*/M_H =1/35$ ($\epsilon\simeq0.17$, light blue), our HMF model with $M_*/M_H =f(M_H)$  (solid green), and the dust-free model from \citet{ferrara2023} (dashed red). \textit{Right:} The observed evolving UV luminosity density, $\rho_{\rm{UV}}$, and cosmic star-formation rate density, $\rho_{\rm{SFR}}$, at $z>8$ compared to the predictions of our model of the HMF in which the typical stellar population age is allowed to be redshift-dependent (solid green line). The dashed green line shows the extreme-redshift prediction of our HMF model if the stellar population age is simply fixed to $10\, \rm Myr$ at $z\geq12.5$. The two {\it inset} panels show the age (left) and inferred formation redshift (right) of the stellar populations required to best fit our evolving HMF model to the UV LF at each redshift. }  
    \label{fig:rho_uv_model}
\end{figure*}

To explore the physical processes which might explain our observational measurements we have constructed a simple model of galaxy evolution based on the evolving dark matter halo mass function (HMF). We first calculated the evolving HMF at $z=8-15$ using \texttt{HMFcalc} \citep{murray2013} with the model from \citet{reed2007}. This was then converted to an evolving galaxy stellar mass function (GSMF) simply using a form of the mass-dependent stellar-mass to halo-mass ratio consistent with that at $z\simeq0$ \citep{behroozi2010}. This step essentially applies the impact of the inferred feedback processes at both high and low halo masses which regulate the shape of the GSMF and, correctly or incorrectly, assumes this is unchanged with redshift (see Appendix \ref{sec:appendix}). The derived evolving GSMF  was then converted to a UV LF at each redshift by a scaling equivalent to assuming that a galaxy with a stellar mass of $\log(M_*/{\rm M_{\odot})=9}$ has a UV luminosity equivalent to $M_{\rm UV}=-22.47$. This mass to UV magnitude conversion was determined from a BC03 stellar population model \citep{bruzual2003} with a metallicity of $Z/{\rm Z_{\odot}=0.2}$, a \citet{chabrier2003} IMF and an assumed constant star-formation history with an age of 30\,Myr. We compare the predictions of this simple model to the observed evolution of $\rho_{\rm{UV}}$ in the right-hand panel of Fig.~\ref{fig:rho_uv}, and show a detailed comparison of the LF predicted by this model with our observed UV LF at $z=11$ in the left-hand panel of Fig.~\ref{fig:rho_uv_model}. 

We show our primary model by the dashed green line which is consistent with the data at $z\geq10$ but overshoots the observations at $z<10$. To attempt to correct this we introduce a dust component to the model by adding mass-dependent UV dust attenuation, $A_{1500}$, given by the dust to stellar-mass relation at $z=2$ derived by \citet{mclure2018}. This model is shown by the orange line in the right-hand panel of Fig.~\ref{fig:rho_uv} which lies slightly below the green dashed line, but still overshoots the data at $z<10$. This is perhaps unsurprising as galaxies at $z<10$ are likely to have average ages older than the assumed value of 30 Myr. Moreover, additional feedback at low stellar masses in the reionization era, or reduced halo occupation could also play a role in reducing the predicted $\rho_{\rm UV}$ at $z<10$.

We also show two secondary models. The first is created by altering the step where we apply feedback at the low- and high-mass ends of the HMF. Instead we simply use $M_*/M_H=1/35$, which is the peak value found in the present-day Universe, corresponding to the observed maximum in historical star-formation efficiency ($\epsilon\simeq0.17$). This effectively provides an upper limit on what is physically permitted by the assumed $\Lambda$CDM cosmology while assuming that star-formation efficiency has never been more efficient at any mass than for present-day Milky-Way mass galaxies. This simple removal of both high- and low-mass feedback predicts the evolution of $\rho_{\rm{UV}}$ shown by the dashed light-blue line in the right-hand panel of Fig.~\ref{fig:rho_uv} and the form of the $z=11$ UV LF shown by the solid light-blue line in the left-hand panel of Fig.~\ref{fig:rho_uv_model}. 

Clearly the predictions of this  modified model vastly  exceed the observational results, demonstrating that our new measurements of the UV galaxy LF certainly do not threaten the viability of the standard $\Lambda$CDM cosmological model. We also show another modified model illustrated by the dashed dark-blue line in the right-hand panel of Fig.~\ref{fig:rho_uv}. This is the same as the primary model except that it assumes a different mapping from galaxy stellar mass to UV luminosity. Here we assume that a galaxy of stellar mass of $\log(M_*/{\rm M_{\odot}})=10$ has a UV luminosity equivalent to  $M_{\rm UV}=-22.4$ as assumed at $z=7$ in \citet{bowler2014}. This demonstrates that a much steeper decline is inevitably predicted when fixing the stellar mass to UV luminosity mapping at $z=7$ assuming no dust obscuration.

\subsubsection{The effect of dust at $z>10$}
As mentioned above, \citet{ferrara2023} have presented a dust-free model of the UV LF and suggest physical mechanisms by which the dust could be expelled from galaxies at $z>10$. The galaxy UV LF predicted by this model at $z = 11$ is shown by the dashed red line in the left-hand panel of Fig.~\ref{fig:rho_uv_model}, where it can be seen to be in excellent agreement with our observational data points.
However, as shown in the right-hand panel of Fig.~\ref{fig:rho_uv}, there is no significant difference in the inferred evolution of $\rho_{\rm{UV}}$ predicted by models which do or do not include dust obscuration at $z>10$. This is because the dust-mass relation deduced by \citet{mclure2018} at $z=2$ indicates that there is essentially no significant dust attenuation ({\it i.e.} $A_{1500} \simeq 0$) for stellar masses below $\log(M_*/{\rm M_{\odot}})\simeq8.4$, and, based on our primary model, the vast majority of the galaxies currently observed at $z>10$ have stellar masses smaller than this. Indeed, our brightest bin at $z=11$, corresponding to $M_{\rm UV}=-21.25$,  equates to a stellar mass of $\log(M_*/{\rm M_{\odot}})=8.5$ according to our adopted mass to UV luminosity conversion. In other words, even assuming a Universe as dusty as observed at $z \simeq 2$ ({\it i.e.} at "cosmic noon", where dust-obscured star formation dominates $\rho_{\rm{SFR}}$) we would not expect any of the galaxies that contribute to our observed UV LF at $z \simeq 11$ to be significantly attenuated by dust in the rest-frame UV. Or, in other words, the stellar-mass range (and hence $M_{\rm UV}$ regime) in which dust, if present at such early times, would have an observable impact has not yet been accessed in this study. Thus, within the range of UV luminosities probed by the {\it JWST} imaging surveys analysed here, we do not need to invoke any physical mechanism to remove or destroy dust to explain the data, as the observed galaxies are not expected to be massive enough to contain significant quantities of dust at any redshift. Further work, including the exploitation of still larger-area surveys, is thus required to measure the very bright end of the galaxy UV LF ($M_{\rm UV}\lesssim-21.5$) at $z=11$, and  hence to determine whether dust removal processes are required to explain the properties of higher-mass galaxies at these early times. 

Despite this current lack of robust statistical constraints on the bright-end form of the $z \simeq 11$ galaxy UV LF at  $M_{\rm UV}<-21.5$, a few `bright' galaxy candidates have been uncovered in this luminosity regime at $z > 10$. These include the spectroscopically-confirmed $M_{\rm UV}\simeq-21.8$ galaxy at $z=10.6$ reported by \citet{bunker2023b} and \citet{tacchella2023}. Interestingly, these authors report a modest dust attenuation of $A_{\rm V}=0.17$ for a stellar mass of $\log(M_*/\rm M_{\odot})=8.73$, consistent with the stellar mass-to-UV magnitude mapping and dust-mass relation used in our primary model. In addition we note that \citet{casey2023} have reported an initial sample of more luminous ($M_{\rm UV}\lesssim-21$) galaxy candidates from the first half of the COSMOS-Web programme \citep{caset2023b}, derived  from imaging covering $\sim0.28\, \rm deg^2$. This yields an estimate of the LF at $z=11$ indicated by the dark-green data-points in the bottom-left panel of Fig.~\ref{fig:LF}. These inferred number densities are consistent with our best-fitting DPL and suggests that there may indeed be galaxies massive enough to require dust-removal mechanisms at these extreme redshifts. However, the lack of contiguous filter coverage in the COSMOS-Web programme may result in more significant numbers of low-redshift contaminants in the high-redshift galaxy samples, in turn leading to more uncertain and potentially biased estimates of the bright end of the galaxy UV LF at extreme redshifts. Spectroscopic verification of these luminous high-redshift candidates is therefore required to accurately constrain their abundance. 

\subsubsection{An age-dependent model}
Finally, we alter our primary model to introduce a stellar population age-dependence which creates a redshift-dependent mapping of stellar mass to UV luminosity. This was determined by adjusting the typical stellar age at a given redshift to best map the GSMF onto the UV LF at $z=8-14.5$. This model is shown by the green solid line in the right-hand panel of Fig.~\ref{fig:rho_uv_model}. Unsurprisingly, and largely by design, this provides an excellent representation of the evolution of $\rho_{\rm{UV}}$ out to $z \simeq 13$. Also unsurprising, and physically sensible, are the relatively young inferred ages of the stellar populations at each redshift which dominate the rest-frame UV light, as tabulated in Table~\ref{tab:age_model}. 

The required stellar ages are consistent with those obtained from fitting SED models to the \textit{JWST}/NIRCam photometry. For example, \citet{robertson2022,robertson2023b} derive stellar ages of $t_*\simeq10-70$\,Myr for galaxies at $z\geq10$. This trend of younger stellar ages at increasing redshift is also consistent with theoretical predictions, where increased gas accretion rates at high redshift lead to increased star-formation rates for fixed stellar mass \citep{mason2015,mason2022}. Therefore at fixed stellar mass, galaxies at higher redshifts have greater UV luminosities and younger stellar ages, consistent with the results of our model. What is more surprising, and potentially very interesting, is the extent to which, for $z =$ 8, 9, 10, and 11, the required stellar population ages converge on a common formation time corresponding to $\simeq 380$\,Myr after the Big Bang (equivalent to a formation redshift $z_f \simeq 12$). This is again tabulated in Table~\ref{tab:age_model}, with the results of this analysis shown in the two inset panels in the right-hand panel of Fig.~\ref{fig:rho_uv_model}. We note that our fit quality and overall conclusion is unchanged with different choices of the HMF in the model, with only a modest increase in the required stellar ages resulting from the adoption of a \citet{sheth1999} HMF (pushing the galaxy emergence epoch back slightly to $z\simeq12.5$) or decrease for a \citet{tinker2008} HMF. Our adopted model from \citet{reed2007} is positioned in the middle of the scatter between different models of the HMF at $z>8.5$.

\begin{table}
	\centering
	\caption{The age-dependent UV magnitude mapping to stellar mass as a function of redshift for our age-dependent model. This model is based on a BC03 stellar population model \citep{bruzual2003} with a metallicity of $Z/{\rm Z_{\odot}=0.2}$. The first column is the redshift. The second column is the UV magnitude, $M_{\mathrm{UV}}$, that is mapped to a stellar mass of $\log(M_*/{\rm M_{\odot}})=9$ determined by the age given in the third column. The fourth column is the formation time after the Big Bang associated with this age and the final column is the formation redshift.}
	\label{tab:age_model}
    \setlength{\tabcolsep}{4pt} 
	\renewcommand{\arraystretch}{1.15} 
	\begin{tabular}{lcccc} 
		\hline
		z & $M_{\mathrm{UV}}$ /AB mag & Age /Myr & $t_{\mathrm{form}}$ /Myr & $z_{\mathrm{form}}$\\
		\hline
  8 & $-20.68$ & 178 & 451 & 10.23\\
  9 & $-20.92$ & 151 & 386 & 11.46\\
  10 & $-21.65$ & 71 & 395 & 11.28\\
  11 & $-22.47$ & 29 & 380 & 11.60\\
  12.5 & $-23.33$ & 7 & 336 & 12.69\\
  14.5 & $-24.18$ & 3 & 276 & 14.61\\
  \hline
	\end{tabular}
\end{table}

Obviously the small number of galaxies discovered at $z \geq 13$ require a still higher formation redshift (albeit with now very young stellar ages, corresponding to a formation redshift $z_f \leq 14$), but there is nothing in our analysis which would {\it a priori} have required an inferred common epoch of formation for the galaxy populations observed over the redshift range $z \simeq 8 - 11$. These results indicate not only that the observed high-redshift evolution of the UV galaxy LF (and hence $\rho_{\rm{SFR}}$) can be explained without requiring any changes to cosmology, star-formation efficiency, or indeed dust, but intriguingly they also point towards the rapid emergence of early galaxies at $z \simeq 12-13$, consistent with the first suggestions of a steeper decline in galaxy number density at $z \geq 13$ seen here in Fig.~\ref{fig:rho_uv}.

In this context, to illustrate what is {\it expected} to happen once ever-younger stellar populations can no longer offset the rapid decline of the halo mass function back to earlier times, we also plot a green dashed line in the right-hand panel of Fig.~\ref{fig:rho_uv_model} which shows the extreme-redshift prediction of our model when the stellar population age is simply fixed to $10\, \rm Myr$ at $z\geq12.5$. The prediction is that, unless the inevitable extreme-redshift decline in the halo mass function is offset by more extreme/exotic stellar populations or enhanced star-formation efficiency, UV luminosity density is expected to decline more rapidly, by roughly two orders-of-magnitude from $z \simeq 13$ to $z \simeq 16$. 

\section{Conclusions}
\label{sec:conclusions}
We have completed an analysis of the major Cycle-1 \textit{JWST}/NIRCam imaging surveys PRIMER, JADES and NGDEEP, covering a total area of $\simeq 370$ sq. arcmin and reaching a 5-$\sigma$ depth of $\simeq 30$ AB mag in the deepest regions. Rather than simply selecting galaxy candidates at high redshift by their "best" photometric redshift, we selected all galaxies that have at least a 5 per cent probability of lying at $z\geq8.5$ and consider their $p(z)$. Through careful selection of galaxies we have derived the $p(z)$ for 2548 galaxies using a UV LF prior and hence calculated the evolution of the galaxy UV LF at $8.5<z<15.5$. Our multi-field approach has allowed new constraints to be placed on the form of the UV LF spanning a UV luminosity range corresponding to $\simeq 4$ AB mag over the redshift range  $z=9-12.5$. This has allowed us to reach a number of conclusions.

First, the large dynamic range in UV luminosity resulting from our multi-field approach has enabled us to define the shape of the UV LF at $z = 9 - 12.5$ from $M_{\rm UV}\simeq-21$ to $M_{\rm UV}\simeq-17$. We have fitted our new measurements with a double-power law (DPL) functional form and explored how the parameters evolve with redshift. We find a lack of evolution in the bright- and faint-end slopes as well as at most only modest evolution in the characteristic magnitude, $M^*$. Much stronger evolution is seen in the LF normalisation suggesting that the evolution of the LF at $z=9-12.5$ is dominated by density evolution rather than luminosity evolution. 

Second, our new measurements of the UV LF have yielded improved constraints on the evolution of cosmic star-formation rate density, $\rho_{\rm{SFR}}$, at $z\geq9$. We find good agreement with prior work showing a smooth, slow evolution over the redshift range $z=9-12.5$ which can be described with a log-linear relationship. Indeed, in the present study we see very little evolution between $z=9$ and $z=11$. We have also presented a tentative measurement of the galaxy number density at $z=14.5$ which suggests the onset of a steeper decline in $\rho_{\rm{SFR}}$, below the extrapolation of the log-linear relation to these extreme redshifts. This, albeit still necessarily uncertain result, hints at a more rapid build-up of galaxies at very early times corresponding to $z>13$.

Finally, we explore the evolution in the UV LF and $\rho_{\rm{SFR}}$ at $z~\geq~8$ through simple modelling based on the halo mass function. We demonstrate that our measurements are fully consistent with a $\rm{\Lambda}$CDM cosmology, and moreover currently require no change to the star-formation efficiency or the dust properties of galaxies as observed in the low-redshift Universe. Rather, we show that a simple evolution in the ages of the stellar populations can explain our measurements, with (unsurprisingly) the typical age of the galaxies which populate the UV LF decreasing with increasing redshift. Intriguingly we find that the typical ages required at $z \simeq$ 8, 9, 10, and 11 all converge on a time $\simeq 380-330$\,Myr after the Big Bang, equivalent to a formation redshift $z \simeq 12 - 13$. This is consistent with the aforementioned first signs of a steeper drop-off in the galaxy population we find beyond $z \simeq 13$, as expected given the very rapid evolution of the halo mass function at earlier times. 

\section*{Acknowledgements}

We thank Andrea Ferrara for providing his model data. C. T. Donnan, D. J. McLeod, R. J. McLure, J. S. Dunlop, R. Begley, M. L. Hamadouche, F. Liu acknowledge the support of the Science and Technology Facilities Council. J. S. Dunlop also acknowledges the support of the Royal Society through a Royal Society Research Professorship.
A.C. Carnall thanks the Leverhulme Trust for their support via the Leverhulme Early Career Fellowship scheme.
F. Cullen, K. Z. Arellano-C\'{o}rdova and T. M. Stanton acknowledge support from a UKRI Frontier Research Guarantee Grant [grant reference EP/X021025/1]. P. Santini acknowledges INAF Mini Grant 2022 “The evolution of passive galaxies through cosmic time.” R.\,A.\,A. Bowler acknowledges support from an STFC Ernest Rutherford Fellowship [grant number ST/T003596/1]. This work is based [in part] on observations made with the NASA/ESA/CSA James Webb Space Telescope. The data were obtained from the Mikulski Archive for Space Telescopes at the Space Telescope Science Institute, which is operated by the Association of Universities for Research in Astronomy, Inc., under NASA contract NAS 5-03127 for JWST.
These observations are associated with program 1837, 1180, 1210, 2079. CM acknowledges support by the VILLUM FONDEN under grant 37459 and the Carlsberg Foundation under grant CF22-1322. The Cosmic Dawn Center (DAWN) is funded by the Danish National Research Foundation under grant DNRF140.

\section*{Data Availability}

All JWST and HST data products are available via the Mikulski Archive for Space Telescopes (\url{https://mast.stsci.edu}). Additional data products are available from the authors upon reasonable request.



\bibliographystyle{mnras}
\bibliography{primer_LF} 




\appendix

\section{The star-formation efficiency model}
\label{sec:appendix}
\begin{figure}
	\includegraphics[width=\columnwidth]{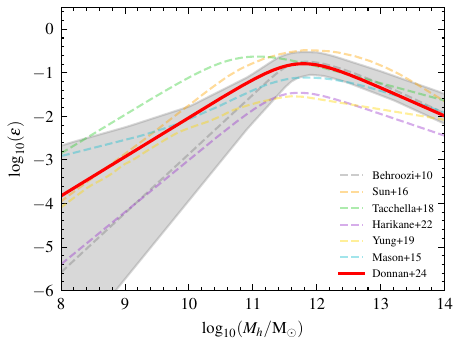}
    \caption{The adopted (redshift independent) star-formation efficiency ($\epsilon$) as a function of halo mass in our model (solid red line) compared to the corresponding relations in the models from \citet{tacchella2018}, \citet{harikane2022}, \citet{yung2019}, \citet{mason2015} and \citet{sun2016}. The dashed grey line shows the $z~=~0.1$ \citet{behroozi2010} stellar-to-halo mass relation (SHMR) which is identical to the relation adopted here at high masses. The low-mass form of our adopted relation is chosen to better track the observational constraints on the SHMR at $z=0$ \citep{wechsler2018}, the uncertainty in which is shown here by the gray shaded region.}  
    \label{fig:SHMR}
\end{figure}

In this short Appendix we detail and explore the role of star-formation efficiency within our simple theoretical model of the evolving galaxy UV LF. As described in Section~\ref{sec:modelling}, we model the star-formation efficiency as a function of halo-mass, $\epsilon(M_{\rm h})$ through the stellar-to-halo mass relation (SHMR), where the stellar and halo masses are simply related by the universal baryon fraction, $f_{\rm b}=0.167$, through the relation

\begin{equation}
    \frac{M_{*}}{M_{h}} = \epsilon(M_{\rm h}) f_{\rm b}
\end{equation}

\noindent
which assumes that the efficiency is solely dependent on the mass of a galaxy's host dark-matter halo. We adopt a functional form for $\epsilon(M_{\rm h})$ given by the double power-law relationship described in \citet{tacchella2018}, namely:

\begin{equation}
    \epsilon(M_{\rm h}) = 2\epsilon_0 \left[ \left(\frac{M_{\rm h}}{M_{\rm c}}\right)^{-\beta} + \left(\frac{M_{\rm h}}{M_{\rm c}}\right)^{\gamma} \right]^{-1}
\end{equation}

\noindent
where $\epsilon_0$ is the peak efficiency, $M_{\rm c}$ is the characteristic mass (the mass at $\epsilon_0$), $\beta$ is the low-mass slope and $\gamma$ is the high-mass slope. We use $(\epsilon_0$, $M_{\rm c}$, $\beta$, $\gamma) = (0.16, 10^{11.7}, 0.9, 0.65)$, and our resulting model relation is shown by the solid red line in Fig.~\ref{fig:SHMR}. Also shown in Fig.~\ref{fig:SHMR} are the observational constraints on the $z \simeq 0$ SHMR as compiled by \citet{wechsler2018} (with the uncertainty captured by the gray shaded region), as well as the $z=0.1$ SHMR from \citet{behroozi2010}. At high masses our model relation is essentially identical to that adopted by \citet{behroozi2010}, but the low-mass form of our model relation is chosen to better track the observational constraints on the SHMR at $z=0$. 

Given that there exist a number of theoretical models of the high-$z$ UV LF, we also compare our adopted relation to the predictions of $\epsilon(M_{\rm h})$ from a number of alternative models. It can be seen that these cover quite a large range, with some adopted  relations clearly inconsistent with the low-redshift observations. For example,  \citet{tacchella2018} assume a higher efficiency ($\sim 1\, \rm{dex}$ across the mass ranges relevant for the $z\geq9$ UV LF) whereas the efficiencies from \citet{mason2015} and \citet{harikane2022} are substantially lower. The semi-analytical model from \citet{yung2019} has a redshift-dependent $\epsilon(M_{\rm h})$ which, for simplicity, we plot here assuming $z=10$, where it transpires to be very similar to our (redshift independent) model at low halo masses. However, one common feature of the models presented by \citet{yung2019,harikane2022} is that they assume a significantly lower star-formation efficiency at the high-mass end ($\sim 1-2\, \rm{dex}$ at $\log(M_{\rm h})>11$) compared either to our own model or to the observationally-defined relationship at $z=0$.

It should be noted that there is an obvious degeneracy between the star-formation efficiency and the age of the stellar population when converting between stellar mass and $M_{\rm UV}$. If a higher efficiency is assumed then the same UV LF can be achieved by an older stellar population. This may explain why although \citet{tacchella2018} assume a significantly higher efficiency than that at $z=0$, they predict lower number densities in the UV LF at $z\geq9$ compared to our model. 

\begin{figure*}
	\includegraphics[width=\textwidth]{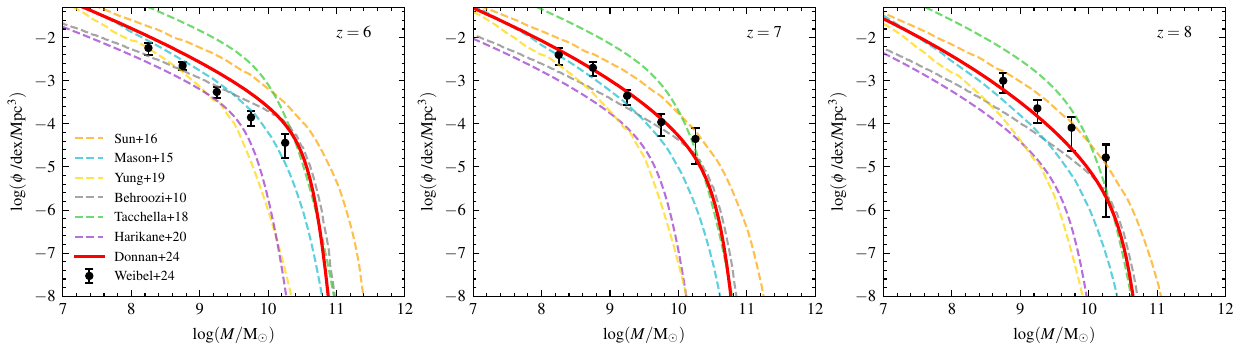}
    \caption{Our model prediction for the evolving galaxy stellar mass function (SMF) at $z=6-8$ (using the $\epsilon(M_h)$ from Fig.~\ref{fig:SHMR}) is shown in each panel by the solid red line where it is compared with the latest {\it JWST} PRIMER-based observational measurements of the SMF from \citet{weibel2024} (black data points). The predictions of the SMF using the alternative $\epsilon(M_h)$ relations shown in  Fig.~\ref{fig:SHMR} from \citet{tacchella2018}, \citet{harikane2022}, \citet{yung2019}, \citet{mason2015}, \citet{sun2016} and \citet{behroozi2010} are shown by the dashed lines for comparison.}  
    \label{fig:SMF}
\end{figure*}

To break this degeneracy requires direct tests of the evolving galaxy stellar mass function predicted by theoretical models. Using the $\epsilon(M_{\rm h})$ the HMF can be converted to a stellar mass function (SMF) by 
\begin{equation}
    M_{\rm *} = \epsilon(M_{\rm h}) f_{\rm b} M_{\rm h}
\end{equation}
where $M_{\rm *}$ is stellar mass. Therefore if one assumes a form for the HMF, the SMF allows a direct measurement of the star-formation efficiency as a function of both stellar and halo mass which is now, thanks to {\it JWST} robustly measurable up to $z\simeq8$. Therefore, as a check on the viability of our model, in Fig.~\ref{fig:SMF} we compare our model prediction for the evolving SMF at $z=6-8$ to the latest observational measurements from \citet{weibel2024}. This comparison shows that our model for $\epsilon(M_{\rm h})$ is able to closely match the measurements of the SMF at $z=6-8$, thus in the process confirming that the star-formation efficiency as a function of halo mass found at $z=0$ is still able to reproduce the SMF out to the highest redshifts yet probed. Fig.~\ref{fig:SMF} also reveals that several of the other theoretical models discussed above, while potentially able to reproduce the high-redshift UV LF, clearly fail this crucial test.

\bsp	
\label{lastpage}
\end{document}